\documentclass{ametsocV6.1}



\usepackage{tabularx}
\usepackage{amssymb}
\usepackage{array}
\usepackage{multirow}
\usepackage{makecell}
\usepackage{lscape}
\newcolumntype{C}{>{\centering\arraybackslash}X}
\renewcommand{\thefigure}{S\arabic{figure}}
\renewcommand{\thetable}{S\arabic{table}}

\title{Improving ensemble extreme precipitation forecasts using generative artificial intelligence}

\authors{
    Yingkai Sha,\aff{a}\correspondingauthor{ksha@ucar.edu}
    Ryan A. Sobash,\aff{a}
    David John Gagne II,\aff{a}}
    
\affiliation{
    \aff{a}{NSF National Center for Atmospheric Research, Boulder, Colorado, USA}}

\abstract{An ensemble post-processing method is developed to improve the probabilistic forecasts of extreme precipitation events across the conterminous United States (CONUS). The method combines a 3-D Vision Transformer (ViT) for bias correction with a Latent Diffusion Model (LDM), a generative Artificial Intelligence (AI) method, to post-process 6-hourly precipitation ensemble forecasts and produce an enlarged generative ensemble that contains spatiotemporally consistent precipitation trajectories. These trajectories are expected to improve the characterization of extreme precipitation events and offer skillful multi-day accumulated and 6-hourly precipitation guidance. The method is tested using the Global Ensemble Forecast System (GEFS) precipitation forecasts out to day 6 and is verified against the Climate-Calibrated Precipitation Analysis (CCPA) data. Verification results indicate that the method generated skillful ensemble members with improved Continuous Ranked Probabilistic Skill Scores (CRPSSs) and Brier Skill Scores (BSSs) over the raw operational GEFS and a multivariate statistical post-processing baseline. It showed skillful and reliable probabilities for events at extreme precipitation thresholds. Explainability studies were further conducted, which revealed the decision-making process of the method and confirmed its effectiveness on ensemble member generation. This work introduces a novel, generative-AI-based approach to address the limitation of small numerical ensembles and the need for larger ensembles to identify extreme precipitation events.}

\begin{document}

%% Necessary!
\maketitle

% Significance Statement (all journals except BAMS)
%
\statement
We use a new Artificial Intelligence (AI) technique to improve extreme precipitation forecasts from a numerical weather prediction ensemble, generating more scenarios that better characterize extreme precipitation events. This AI-generated ensemble improved the accuracy of precipitation forecasts and probabilistic warnings for extreme precipitation events. The study explores AI methods to generate precipitation forecasts and explains the decision-making mechanisms of such AI techniques to prove their effectiveness.

%Requirements: www.ametsoc.org/index.cfm/ams/publications/author-information/significance-statements/

%Introduction
\section{Introduction}\label{sec1}
The accurate prediction of extreme precipitation events is crucial for saving lives and property but continues to challenge our best prediction systems \citep[e.g.][]{sukovich2014extreme,herman2016extreme}. State-of-the-art global numerical weather prediction (NWP) models typically have 10-50 km horizontal grid spacing \citep[e.g.][]{molteni1996ecmwf,buizza2018major,zhou2017performance,zhou2022development}. These horizontal resolutions require parameterization schemes to approximate small-scale processes that contribute to the generation of rainfall, such as convection and the microphysical interactions of cloud and precipitation water particles \citep{stensrud2009parameterization}. Despite the large improvements made over recent decades \citep{bauer2015quiet}, systematic error remains in these parameterization schemes, particularly for the modeling of extreme precipitation events \citep{wilcox2007frequency,wehner2010effect,sun2020improving}. 

Post-processing methods have been proposed for the bias correction and calibration of precipitation forecasts. These methods include both parametric methods, which assume a known predictive distribution, and non-parametric methods, which derive a predictive distribution from the training data with no prior distribution assumptions.
One common parametric method is nonhomogeneous regression \citep{scheuerer2015statistical,baran2016censored}, which is an approach where a regression model predicts the parameters of a parametric distribution, such as a censored, shifted Gamma distribution for precipitation. Bayesian model averaging \citep{sloughter2007probabilistic} estimates a multi-modal predictive distribution from an ensemble of deterministic predictions by learning the predicted variance and weight of each ensemble member. Nonparametric methods include the Analog Ensemble \citep{hamill2006probabilistic,Monache2013-ql}, which identifies the most similar prior NWP predictions to a new prediction and creates a distribution of observed values mapped to those NWP analogs, and quantile regression \citep{bremnes2004probabilistic}, which optimizes regression models to predict conditional predictive quantiles rather than the most likely values. 

While these methods have achieved great success, their primary focus was the overall post-processing quality, which was dominated by the calibration performance of mild-to-moderate precipitation events. Machine-learning-based methods have been introduced to specifically improve extreme precipitation forecasts [e.g. random forest \citep{herman2018money}, feed-forward neural network \citep{bodri2000prediction}, deep neural network \citep{li2022convolutional}]. However, a key challenge of these methods is generating skillful forecast trajectories that represent the evolution of extreme precipitation events. Many machine learning models were trained to predict probabilistic values on locations and forecast lead times independently, whereas end-users may look for multivariate forecast trajectories for flood risk assessments \citep[e.g.][]{lai2020flood,huang2021assessment} and water resource management \citep[e.g.][]{strauch2012using}. 

Producing trajectories that can represent extreme precipitation events properly requires a large ensemble set. Currently, the ensemble size of operational global NWP models is typically limited to 30-50 \citep[e.g.][]{leutbecher2019ensemble,zhou2022development}, which is insufficient to capture extreme events on the very tail side of the precipitation intensity spectra \citep{bevacqua2023advancing}. Most statistical and machine-learning-based post-processing methods cannot solve this problem because they are designed to utilize available ensemble members; they can hardly create new forecast scenarios from an existing ensemble set to improve the estimation of extreme events. One possible solution is producing hundreds of numerical ensemble members from a regional numerical model configuration \citep[e.g.][]{ghazvinian2024deep}, although such efforts are computationally costly.

Recent advances in generative Artificial Intelligence (AI) have brought new insights into extreme weather prediction problems. State-of-the-art generative AI can learn distribution properties from training data and produce conditional samplings from the target distribution \citep{creswell2018generative,bond2021deep,yang2023diffusion}. Compared to physics-based NWP ensembles, generative AI can expand ensemble sizes by producing more forecast members at minimal computational cost. This enlarged generative ensemble set would contain possible evolutions of the state of the atmosphere, thus providing better support for the estimation of high-impact extreme weather.

Several studies have leveraged generative AI in either NWP or AI-based weather prediction. On regional scales, \citet{sha2024generative} found that generated ensemble members from deterministic convection-allowing model forecasts improved probabilistic estimations of tornadoes, hail, and wind gusts. On global scales, \citet{li2024generative} applied generative AI to post-process and emulate ensemble forecasts, resulting in improved forecast skill and more accurate predictions of extreme weather. \citet{zhong2023fuxi} integrated generative AI with an AI weather prediction model to produce multi-day forecasts with finer-scale spatial details. The generated forecasts outperformed the original AI weather forecasts on various extreme-weather-based metrics.

Motivated by the challenge of multivariate extreme precipitation post-processing and the application of generative AI in extreme weather prediction, this research proposes a post-processing framework that incorporates generative AI to improve the estimation of extreme precipitation events. Specifically, we aimed to produce a skillful generative ensemble of 6-hourly precipitation forecasts out to 6 days. This generative ensemble is expected to provide precipitation forecast trajectories that characterize extreme precipitation events properly and can be summarized with probabilistic guidance.

The methodology of this research was developed over the conterminous United States (CONUS) using the Global Ensemble Forecast System (GEFS) as inputs and the Climate-Calibrated Precipitation Analysis (CCPA) as targets. The following research questions are addressed: (1) How can generative AI methods be incorporated into precipitation forecast post-processing, and how well do they verify at producing reliable and discriminative probabilistic forecasts at extreme precipitation thresholds? (2) Can we explain the performance of AI-based precipitation post-processing methods, and what insights can we gain from such explainability analysis? By answering these, the authors examine the effectiveness of generative AI in extreme precipitation forecasts and explore its decision-making mechanisms for post-processing. Broadly, the authors also wish to introduce generative AI to severe-weather-related studies and inspire future creative works.

%Research domain and data
\section{Research domain and data}\label{sec2}

%Region of interest
\subsection{Region of interest and the definition of extreme precipitation events}\label{sec21}

\begin{figure}[t]
\noindent\includegraphics[width=\textwidth]{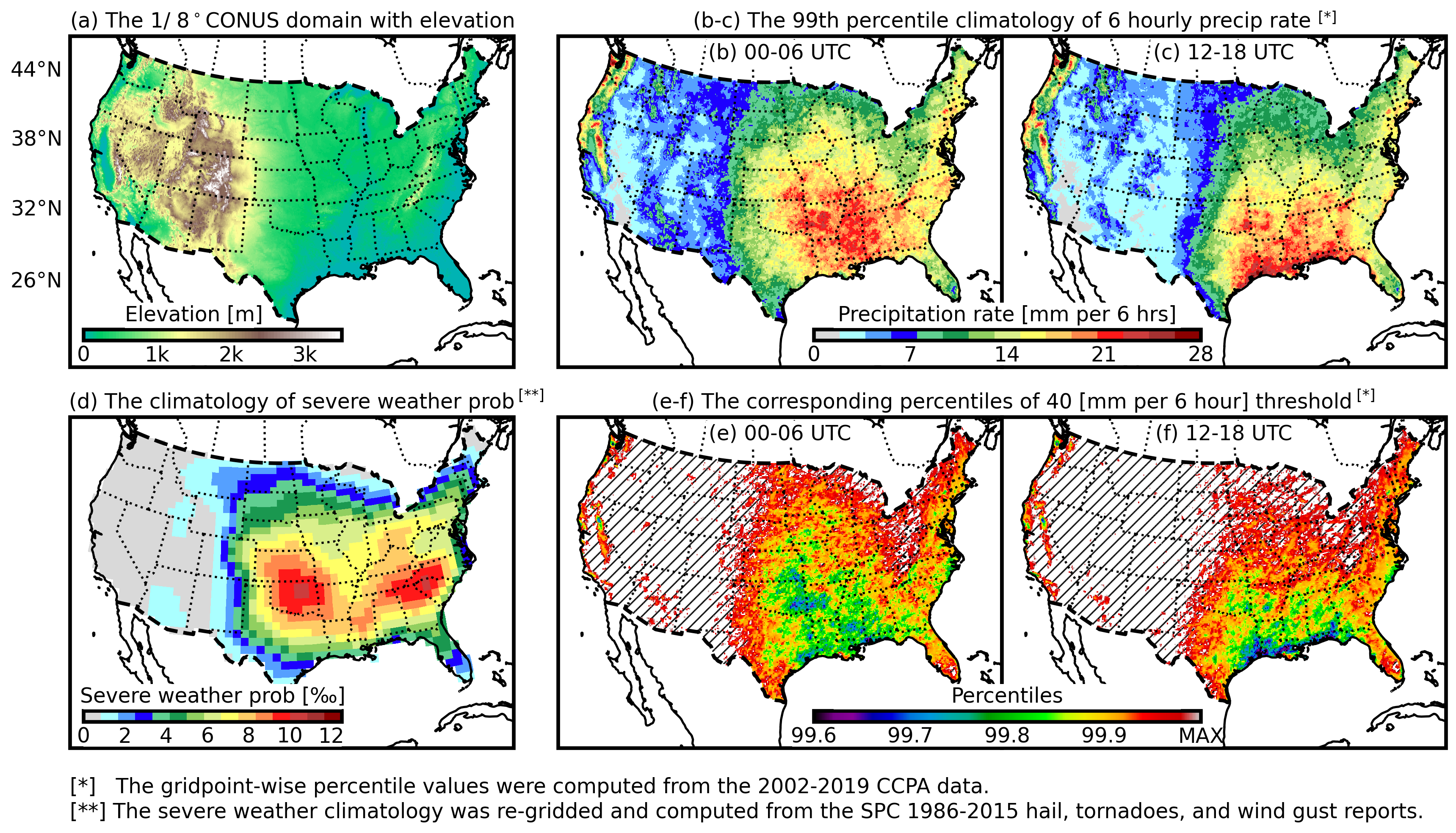}\\
\caption{(a) The 0.125$^\circ$ grid spacing CONUS domain with shaded elevation. (b) and (c), the 2002--2019 climatology of gridpoint-wise 99th percentile values of 6 hourly precipitation rates for 0000--0600 and 1200--1800 UTC, respectively. (d) 1986--2015 climatological probabilities of tornadoes, hail, and wind gusts derived from the National Oceanic and Atmospheric Administration (NOAA) Storm Prediction Center (SPC) reports. (e) and (f), the corresponding 2002--2019 climatological percentiles of 40 mm per 6-hour threshold for 0000--0600 and 1200--1800 UTC, respectively. Hatched area in (e) and (f) means percentiles cannot be estimated as 40 mm per 6-hour is close to or larger than the historical maximum.}\label{fig1}
\end{figure}

This research focuses on precipitation events within the CONUS (Fig.~\ref{fig1}a). Following the definition of the Intergovernmental Panel on Climate Change (IPCC), Sixth Assessment Report (AR6), gridpoint-wise 99th percentile values were used as thresholds to identify extreme precipitation events \citep{portner2022climate}. These values were estimated separately for different time periods of the day (00--06, 06--12, 12--18, and 18--00 UTC) to capture diurnal variations. The 00--06 and 12--18 UTC values were provided as examples in Fig.~\ref{fig1}b and c, respectively. 

The percentile-based peak-over-threshold approach can be inconsistent due to large regional differences \citep{wang2020spatial}. Thus, a fixed precipitation rate threshold of 40 mm per 6 hours was used to identify extreme precipitation events. Fig.~\ref{fig1}e-f provide the corresponding percentiles of the 40 mm per 6 hours threshold for 00--06 and 12--18 UTC. Overall, this fixed threshold is more extreme than the 99th percentile definition; it largely ignores dry areas on the west side of the Rocky Mountains and emphasizes precipitation in the Great Plains and the Southeastern US (cf. Fig.~\ref{fig1}d and e-f). The 00--06 UTC (i.e., evening-to-night local time) patterns in Fig.~\ref{fig1}e are primarily explained by the nocturnal convection over the Great Plains \citep[e.g.][]{jiang2006role,johnson2017,blake2017}, whereas the 12--18 UTC (i.e., morning-to-afternoon local time) patterns in Fig.~\ref{fig1}f are related to summertime deep convection \citep[e.g.][]{tian2005diurnal}, small-scale convection introduced by the sea-breeze circulation \citep[e.g.][]{hill2010summertime}, and tropical cyclones originating from the Gulf of Mexico \citep[e.g.][]{shepherd2007quantifying}. These types of extreme precipitation events are difficult to predict but also have a high impact by causing flash floods \citep[e.g.][]{nair1997numerical,caracena1979mesoanalysis,smith2001extreme}. In addition, other fixed thresholds, ranging from 1 mm to 35 mm per 6 hours, were examined to provide comprehensive views of the performance of post-processing methods.

%Forecast data
\subsection{Forecast data}\label{sec22}

This research aimed to improve the extreme precipitation forecasts from GEFS version 12 (GEFSv12; \citealt{zhou2022development}). The GEFSv12 is a state-of-the-art real-time ensemble forecast system operated by NOAA since September 2020. GEFSv12 implements the Global Fluid Dynamics Laboratory (GFDL) Finite-Volume Cubed-Sphere (FV3) dynamical core, ensemble Kalman filter-based data assimilation to generate initial condition uncertainty, and the GFDL cloud microphysics. Its quantitative precipitation forecasts over CONUS were largely improved from its previous versions \citep{zhou2022development}. GEFSv12 has 0.25$^\circ$ output horizontal grid spacing and 64 vertical hybrid levels. It produces 31-member ensemble forecasts 4 times per day, with 3-hourly output available within the first 10 forecast days. In this research, the 00 UTC GEFSv12 initializations and 6 hourly total precipitation forecast (APCP) were selected as the main variable, whereas total-column precipitable water was used for one of the baseline methods. 

The operational GEFSv12 comes with a 30-year reforecast archive to support post-processing studies \citep{guan2022gefsv12}. This reforecast dataset was produced from the same dynamical core, ensemble generation, and model physics as the operational GEFSv12, but with 5 members and 0000 UTC initializations only. The phase-two, 2000--2019 reforecasts, initialized from the GEFSv12 reanalysis \citep{hamill2022reanalysis}, were used by this research as training data.

%Observations
\subsection{Analysis data}\label{sec23}

CCPA \citep{hou2014climatology} was used in this research to represent the analyzed state of precipitation. CCPA is a precipitation dataset that covers the entire CONUS. It statistically adjusts and combines the National Centers for Environmental Prediction (NCEP), Climate Prediction Center (CPC) unified global daily gauge analysis, and the NCEP Stage IV multi-sensor quantitative precipitation estimation \citep{hou2014climatology}. CCPA was used as the verification target of the operational GEFS system \citep{zhou2017performance,zhou2022development} and has been applied as training and verification targets in various GEFS-based post-processing studies \citep[e.g.][]{scheuerer2015statistical,hamill2018probabilistic,stovern2023improving,hamill2023improving}. The CCPA was also used as a climatology reference, including the estimation of gridpoint-wise precipitation Cumulative Distribution Functions (CDFs), which were used to define percentile-based extreme precipitation events (see Section \ref{sec2}.\ref{sec21}) and compute skill scores.

%Method
\section{Methods}\label{sec3}

\begin{figure}[t]
\noindent\includegraphics[width=\textwidth]{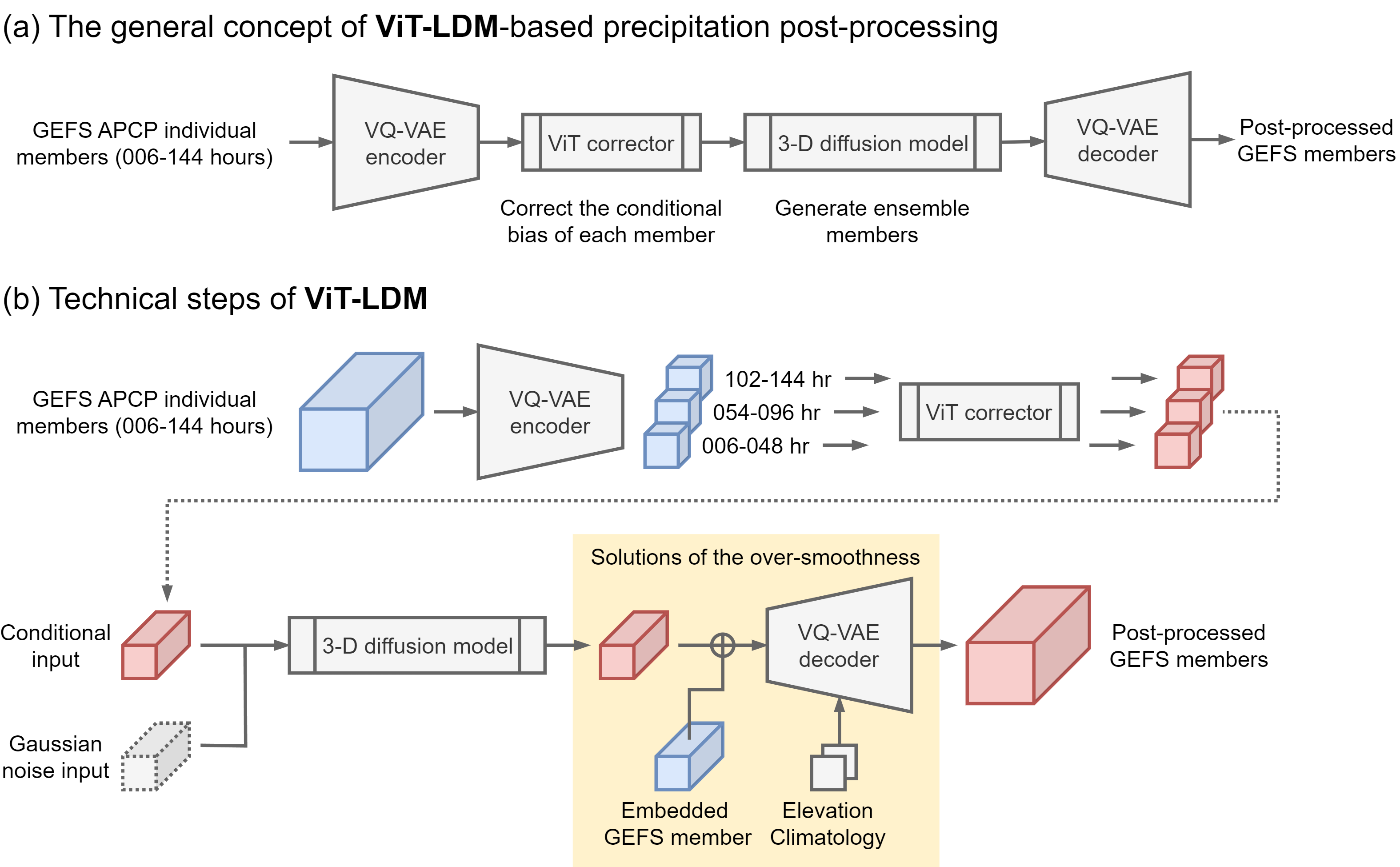}\\
\caption{The general concept (a) and technical steps (b) of ViT-LDM. Steps that solve the over-smoothness problem of the VAE-based latent diffusion are highlighted using a yellow background color.}\label{fig2}
\end{figure}

Two neural-network-based post-processing steps were combined as the main methodology of this research (Fig.~\ref{fig2}). First, a 3-D Vision Transformer (ViT) was proposed to reduce the forecast bias of each GEFSv12 member. Second, each bias-corrected member was used as the conditional input of a diffusion model, which generates post-processed members as outputs. The two steps above were conducted within the latent space created by a Vector-Quantized Variational Autoencoder (VQ-VAE), with the VQ-VAE encoder projecting GEFS members into the latent space and the VQ-VAE decoder projecting the latent space outputs back to the real space. The term ``Latent Diffusion Model (LDM)'' was used to highlight the application of VQ-VAE, and hereafter, the method is named ``ViT-LDM''.
 
The ViT-LDM post-processing was trained from 2002 to 2019 using the 6-hourly GEFS reforecasts out to 6 days (i.e., 06--144 hours lead times) and the CCPA data. The validation set was a 10\% random sampling from the training set and fixed for all training steps. When applied to the operational GEFSv12 in 2021, ViT-LDM generates 2 post-processed members from each GEFSv12 member, thus producing 62 generative members from all 31 operational members. The generated 62 members were verified against the CCPA data from 1 January to 31 December 2021 with a focus on extreme precipitation events. The basics and applications of VQ-VAE, ViT, and LDM are introduced in this section. Their hyperparameter optimization and other related information are summarized in the Appendix.

\subsection{Latent space projection using VQ-VAE}\label{sec31}

\begin{figure}[t]
\noindent\includegraphics[width=\textwidth]{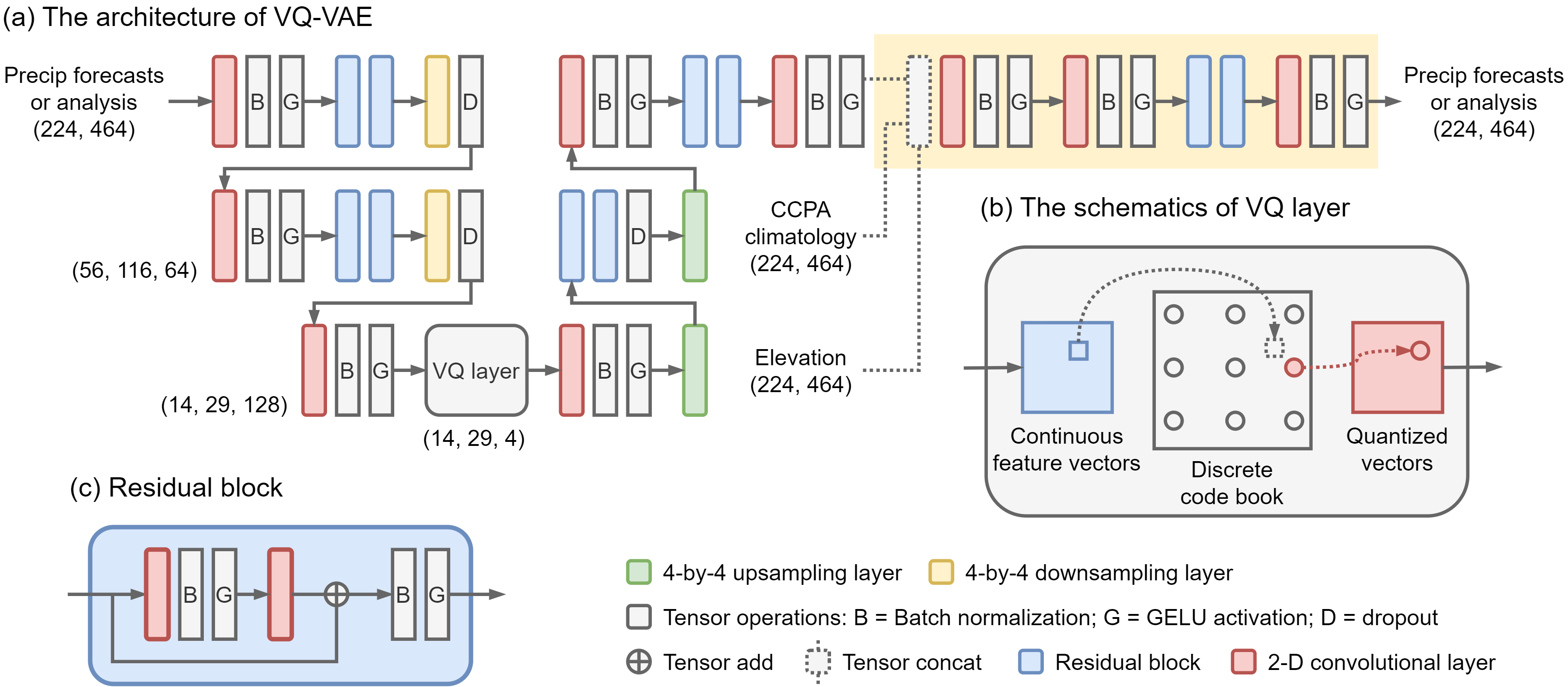}\\
\caption{(a) The architecture of VQ-VAE with 2-D convolutional layers, down- and upsampling layers, batch normalization \citep{ioffe2015batch}, Gaussian Error Linear Unit (GELU;  \citealt{hendrycks2016gaussian}) activation function, and dropout \citep{srivastava2014dropout}. (b) The schematics of the VQ layer and its dashed line arrows represent identical mapping. (c) The design of the residual block. A separate output section is highlighted using a yellow background color.}\label{fig3}
\end{figure}

A VQ-VAE \citep{van2017neural} was employed to convert gridded precipitation fields, either from the GEFS APCP forecasts or the CCPA data, into a compressed and regularized latent space, enabling effective bias correction and ensemble generation. The VQ-VAE of this research was designed based on 2-D convolutional layers. Its encoder contains two 4-by-4 downsampling layers with 16 times compression on latitude and longitude dimensions. The VQ-VAE decoder has two 4-by-4 upsampling layers; it takes the encoded latent space features as input and projects them to the original size with minimum information loss. The output section of the VQ-VAE decoder has an extended sub-structure that accepts elevation and CCPA climatology as additional inputs to improve the decoding quality. The technical highlight of VQ-VAE is its Vector-Quantization (VQ) layer. The VQ layer converts continuous encoded information into discrete values by selecting the closest-distanced vector from a discrete and learnable ``codebook''. VQ-VAE can be viewed as a VAE that produces discrete latent space embeddings.

The use of VQ-VAE in this research brought two major benefits: (1) The VQ-VAE latent space projection reduces data size, so the ViT and LDM can be designed and trained more effectively. (2) A regularized VAE latent space disentangles the input data. This means each VAE latent variable would represent its own factors of variation, and small perturbations within the VAE latent space would not lead to dramatically different outputs. The disentanglement property benefits the stability of ensemble member generation and model interpretation. Compared to direct diffusion, a known disadvantage of VAE-based latent diffusion is the over-smoothness of its outputs \citep[e.g.][]{yang2024lossy}. Two steps were applied to solve this problem: (1) The embedded GEFS APCP forecasts were linearly combined with the generated latent information to guide the VQ-VAE decoder in producing more physically realistic outputs in inference (Fig.~\ref{fig2}); (2) The decoder sub-structure that incorporates elevation and climatology information, as mentioned in the previous paragraph, was also aimed to mitigate the over-smoothness issue (Fig.~\ref{fig2}, Fig.~\ref{fig3}).

VQ-VAE features self-supervised training. Its optimization objective contains three components \citep{van2017neural}:

\begin{equation}\label{equ1}
\mathcal{L}\left(x\right) = \left\|x-z_d\left[z_e\left(x\right)\right]\right\|_2^2 + 
\left\|\text{sg}\left[z_e\left(x\right)\right] - e\right\|_2^2 + 
\beta \left\| z_e\left(x\right) - \text{sg}\left(e\right)\right\|_2^2
\end{equation}

\noindent
Where $x$ is a training batch, $\left\|\cdots\right\|_2^2$ is the mean squared error computation, $z_e$ and $z_d$ are the VQ-VAE encoder and decoder, respectively, $e$ is the codebook, and ``sg'' is the stop-gradient operator, which fixes the target from being updated by the current gradient descent step. The first term of equation (\ref{equ1}) is the reconstruction loss; it minimizes the difference between the input and the reconstructed input. The second term is the codebook loss; it updates the discrete codebook values to keep them close to the continuous encoded information. The last term of the equation (\ref{equ1}) is the commitment loss; it regularizes the encoder to prevent its encoded continuous values from diverging from the current codebook. $\beta$ is a constant hyperparameter that defines the relative importance of commitment loss.

The VQ-VAE described above was trained on the 1/8$^\circ$ CCPA data with (224, 464) input sizes; its decoder produces [(14, 29, 4) sized latent variables as outputs (the last dimension represents hidden-layer channels). The 0.25$^\circ$ GEFS APCP forecasts were linearly interpolated to 1/8$^\circ$ before encoding. The same interpolation and VQ-VAE weights were applied to all forecast lead times and both the GEFS reforecasts and operational forecasts. For data pre-processing, 6 hourly precipitation and the CCPA-based climatology were normalized by using a re-scaled logarithm transformation: $y=\log{\left(0.1x+1\right)}$; the elevation input was normalized to [-1,1] by a linear scaler. Value truncation was applied to the decoder outputs. Precipitation rates lower than 0.1 mm per 6 hours will be replaced by zero. 

\subsection{ViT-based forecasts bias correction}\label{sec32}

\begin{figure}[t]
\noindent\includegraphics[width=\textwidth]{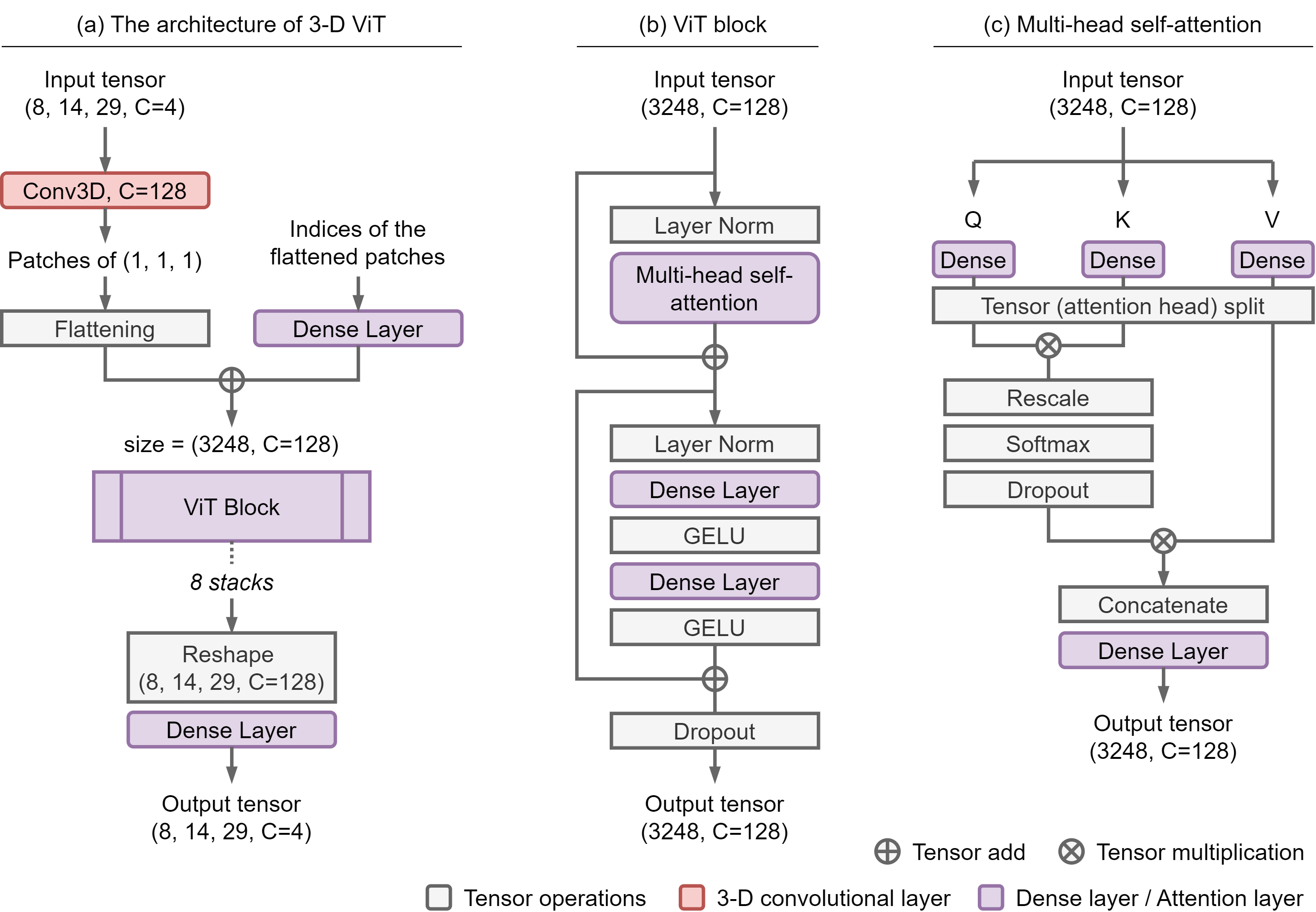}
\caption{(a) The architecture of the 3-D ViT. ``C'' indicates the number of channels. (b) The design of ViT blocks with layer normalization \citep{ba2016layer}, multi-head self-attention \citep{vaswani2017attention}, GELU activation, and dropout. (c) The design of multi-head self-attention. ``Q'', ``K'', and ``V'' represent ``query'', ``key'', and ``value'', respectively, which are three copies of the input tensor for self-attention computation \citep{vaswani2017attention}.}\label{fig4}
\end{figure}

A 3-D ViT \citep{dosovitskiy2020image,vaswani2021scaling,arnab2021vivit} was applied within the VQ-VAE latent space for the bias correction of GEFS APCP forecasts. Its architecture consists of three components: (1) an input section that converts 3-D tensors into embedded patches, (2) stacked ViT blocks that perform attention-based learning, and (3) an output section that converts embedded patches to the original tensor size. The input section conducts patch partition using 3-D convolution kernels, and the positional indices are embedded by a dense layer. This design is similar to many AI-based weather forecast models \citep[e.g.][]{chen2023fuxi}. The ViT block follows the conventional design of \citet{arnab2021vivit}; it features multi-head self-attention to learn the cross-relationships among embedded patches. More advanced ViT designs, such as Shift-window-based Transformers (SwinTs; \citealt{liu2021swin}), were examined during the hyperparameter search, but they did not bring better performance. 

The 3-D ViT operates (1, 1, 1) sized patch partitions with 128 embedded dimensions. Its ViT block has 8 stacks with 4 attention heads. This configuration was trained using the encoded GEFS reforecast ensemble mean as inputs and encoded CCPA as targets. It processes 8 temporal dimensions at once and was trained using the 06-54 hour reforecasts only. This training strategy will be discussed further within the context of AI explainability studies. For inference, the same 3-D ViT was applied to the operational GEFS members on 06--54, 54--102, and 102--144-hour forecasts to generate bias-corrected ensemble trajectories within the VQ-VAE latent space.

%CNN-based end-to-end severe weather prediction
\subsection{Ensemble generation using LDM}\label{sec33}

\begin{figure}[t]
 \noindent\includegraphics[width=\textwidth]{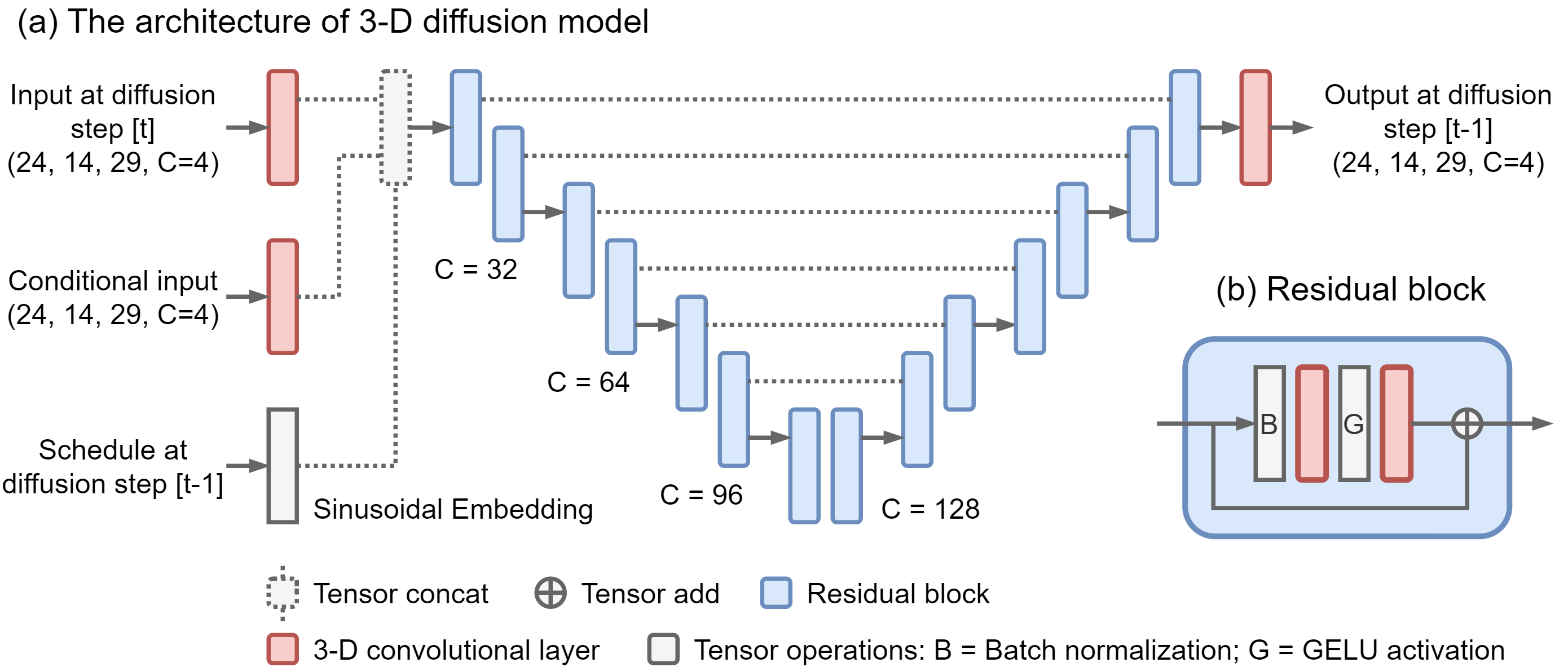}\\
 \caption{(a) The architecture of the 3-D diffusion model. ``C'' indicates the number of channels. (b) The design of residual block}\label{fig5}
\end{figure}

An LDM was implemented to produce generative ensembles conditioned on the bias-corrected GEFS members. The archetype of LDM is the Denoising Diffusion Probabilistic Model (DDPM) proposed by \citet{ho2020denoising}, and it was extended to a 3-D configuration that supports the generation of the entire forecast trajectory.

DDPM contains forward and reverse diffusion processes. For a given sample of the target distribution $X_0\sim q\left(X_0\right)$, the forward diffusion process adds Gaussian noise into the sample iteratively by following a variance schedule:

\begin{equation}\label{equ2}
q\left(X_t|X_{t-1}\right) =  \mathcal{N}\left(X_t; \sqrt{1-b_t}X_{t-1}, b_t\ \mathcal{I}\right)
\end{equation}

\noindent
Where $t = \left\{0, 1, \cdots, T\right\}$ are the diffusion timesteps, $B=\left\{b_0, b_1, \cdots, b_T\right\}$ is the diffusion schedule. The reverse diffusion process is achieved by a neural network $\theta$ that approximates $q_\theta\left(X_{t-1}|X_t\right)$ and recovers the noised sample to its original state $X_0$ iteratively. The optimization objective of DDPM is summarized as follows \citep{ho2020denoising,dhariwal2021diffusion}:

\begin{equation}\label{equ3}
\mathcal{L}\left(t\right) = \left\|\epsilon_{t} - \theta\left(X_t,b_t,V\right)\right\|_2^2
\end{equation}

\noindent
Where $\epsilon_{t}$ is the mean squared error of the predicted accumulated effect of forward diffusion, $V$ is an optional, conditional input that can be incorporated during the reverse diffusion processes to influence the estimation of $q_\theta\left(X_{t-1}|X_t\right)$. For the sample generation of DDPM, $X_T$ is a random draw from $\mathcal{N}\left(0,\mathcal{I}\right)$, and reverse diffused to $X_0$, which results in a generated sample.

The LDM of this research was designed based on the DDPM above and applied within the VQ-VAE latent space. Its architecture is similar to a 3-D Unet \citep[e.g.][]{ronneberger2015u,cciccek20163d} but without down- and upsampling levels. The LDM was configured with a 100-step linear schedule; it takes three inputs (Fig.~\ref{fig5}a): (1) the output of the previous reverse diffusion step, (2) a ViT bias-corrected ensemble member as conditional input, and (3) the diffusion schedule of the current step; it produces the reverse diffusion output on the current step. The LDM was trained using the ViT-corrected reforecasts members as inputs and CCPA as targets. During the sample generation process, the weights of LDM were modified using the exponential moving average. 

\subsection{Baseline methods}\label{sec34}

The combination of Analog Ensemble (AnEn; \citealt{hamill2006probabilistic}) and Ensemble Copula Coupling (ECC; \citealt{schefzik2013uncertainty}) was considered as the baseline of this research (hereafter ``AnEn-ECC''). AnEn is a regression-based method that performs univariate bias correction and ensemble calibration. For each forecast lead time and location, AnEn identifies similar historical dates/times within its reforecast training set and forms an ensemble composed of the CCPA training target at the identified date/times. As a nonparametric method, AnEn leverages a large reforecast archive without requiring an a priori distribution assumption; it is easy to implement and can produce realizations with flexible ensemble sizes. These strengths make AnEn a good option for precipitation forecast post-processing. The AnEn baseline here follows its improved version as introduced by \citet{hamill2015analog} but without supplemental locations. It was trained using the 2002--2019 GEFS reforecasts and the CCPA target.

ECC is a multivariate, nonparametric method that recovers spatiotemporal consistencies from univariate post-processing outputs. Given calibrated AnEn members, ECC applies 31 operational GEFS members as ``dependence templates'' and re-indexes 31 AnEn members based on the rank structure of the selected templates. 

More advanced precipitation post-processing methods were considered for use as a baseline, such as \citet{scheuerer2015statistical} and \citet{stovern2023improving}, but these methods typically produce probabilistic values directly rather than forecasted trajectories with physics-based units. We prefer AnEn-ECC because, similar to ViT-LDM, AnEn-ECC can post-process GEFS precipitation ensembles into forecast trajectories, which allows the flexibility of extreme precipitation verification with different definitions and thresholds (see Section \ref{sec2}.\ref{sec21}).

The original 31 operational GEFS members (hereafter ``GEFS-Raw'') were also used as a baseline. The two baselines, AnEn-ECC and GEFS-Raw, will be contrasted with ViT-LDM in extreme precipitation verification. Note that each of the two baselines contains 31 members, whereas the ViT-LDM generates 62 members. Although the total number of ensemble members is unequal, we think such a comparison is still fair because generating more ensemble members is part of the methodology and purpose of ViT-LDM.

%%%%%%%%%%%%%%%%%%%%%%%%%%%%%%%%%%

%Verification methods
\subsection{Verification methods}\label{sec35}

ViT-LDM and the two baselines were verified from 1 January to 31 December 2021. The general post-processing performance of all methods was examined using the Continuous Ranked Probability Scores (CRPS) and CRP Skill Scores (CRPSS), whereas the performance of extreme precipitation forecasts was verified using the Brier Score (BS) and Brier Skill Score (BSS; \citealt{murphy1973new}). The climatology reference of CRPSS and BSS was derived from the 2002--2019 CCPA data (Section \ref{sec2}.\ref{sec22}).
 
The computation of spatiotemporally aggregated BSSs follows \citet{hamill2006measuring}, with the BS on individual gridpoint and forecast lead times being computed and aggregated first and then converted to BSS by applying the climatology reference. The three-component decomposition of BS and reliability diagrams were also computed to attribute the BSS difference; their computation follows \citet{murphy1973new} and \citet{hsu1986attributes}. Bootstrapping was applied to estimate the confidence intervals of skill scores. It was conducted separately on positive (i.e., extreme precipitation cases) and negative samples to preserve their relative ratios. Two-sided Wilcoxon signed-rank tests were applied to the CRPSS comparisons to determine if skill scores were statistically significantly different.

% Result
\section{Results}\label{sec4}
% Case-based assessments
\subsection{Case-based assessments}\label{sec41}

\begin{figure}[t]
\noindent\includegraphics[width=\textwidth]{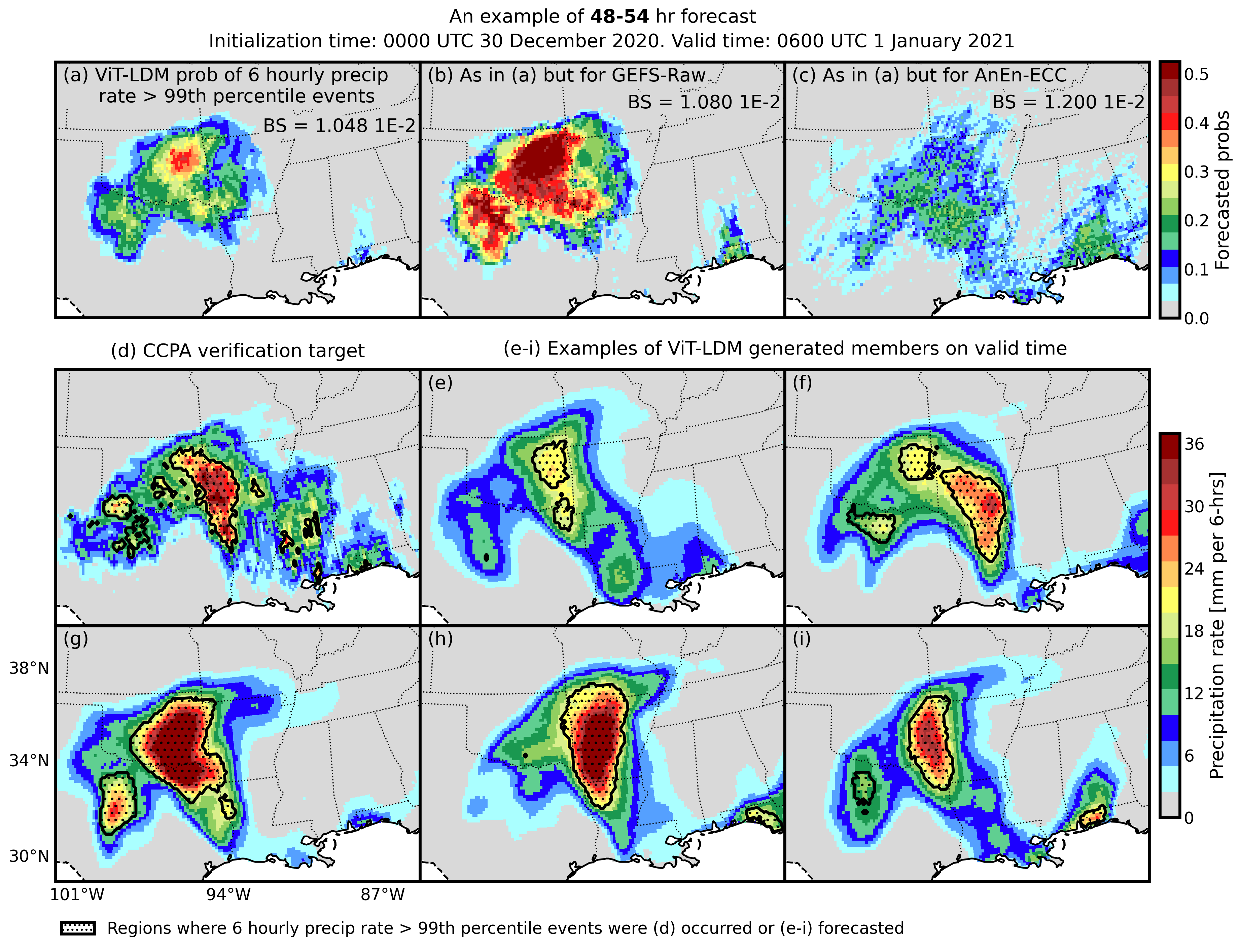}\\
\caption{An example of 48-54 hour forecasts with extreme precipitation events on 0000-0600 UTC 1 January 2021. (a) calibrated probabilities of precipitation rate $>$ gridpoint-wise 99th percentile events. (b) As in (a) but for GEFS-Raw. (c) As in (a) but for AnEn-ECC. (d) the CCPA verification target. (e-i) example of generative ensembles produced by ViT-LDM. Hatched areas represent where the 99th extreme events were analyzed (d) or forecasted (e-i)}\label{fig6}
\end{figure}

A case-based assessment is presented to demonstrate the generative ensemble produced by ViT-LDM. In Fig.~\ref{fig6}, an example of 48-54 hour GEFS forecasts, initialized on 0000 UTC 30 December 2020, is presented. At this time, a synoptic scale system was forecast in the Southeastern US. The system caused extreme precipitation (see the dotted area in Fig.~\ref{fig6}d) and gradually moved towards the East Coast. Part of the ViT-LDM generative ensemble members are shown in Fig.~\ref{fig6}e-i with the corresponding extreme precipitation events highlighted by the dotted area. Comparing the ViT-LDM outputs with the two baselines, several performance highlights are evident:

\begin{enumerate}
    \item Precipitation patterns generated by ViT-LDM shared roughly the same locations as the CCPA verification target. The LDM-based conditional sampling from the bias-corrected GEFS ensemble members has the ability to preserve the broad-scale structure of the forecast event. This ensures that the generative ensemble would not exhibit large spatial discrepancies and place negative impacts on the prediction of extreme precipitation events.
    
    \item Differences in terms of the shape and intensity of the generated precipitation patterns can be found. For example, in Fig.~\ref{fig6}f, the generated precipitation pattern had similar shapes to the CCPA target, but its forecast extreme precipitation area was shifted to the east. In Fig.~\ref{fig6}g and i, extreme precipitation events were forecast on the correct grid points, but the precipitation pattern was extended to the south. Such small-scale variations provided good horizons on how this extreme precipitation event would develop. The probabilistic forecasts, collectively summarized from the ViT-LDM generative ensemble, showed good BS and outperformed the two baselines.

    \item The generated members were smoother than the CCPA verification target. This indicates that, although the ViT-LDM was trained using the 1/8$^\circ$, they may not have the full ability to downscale 0.25$^\circ$ GEFS inputs into the 1/8$^\circ$ target resolution. That said, spatial downscaling is not the purpose of ViT-LDM.
\end{enumerate}

\subsection{General post-processing performance verification}\label{sec42}

\begin{figure}[t]
\noindent\includegraphics[width=\textwidth]{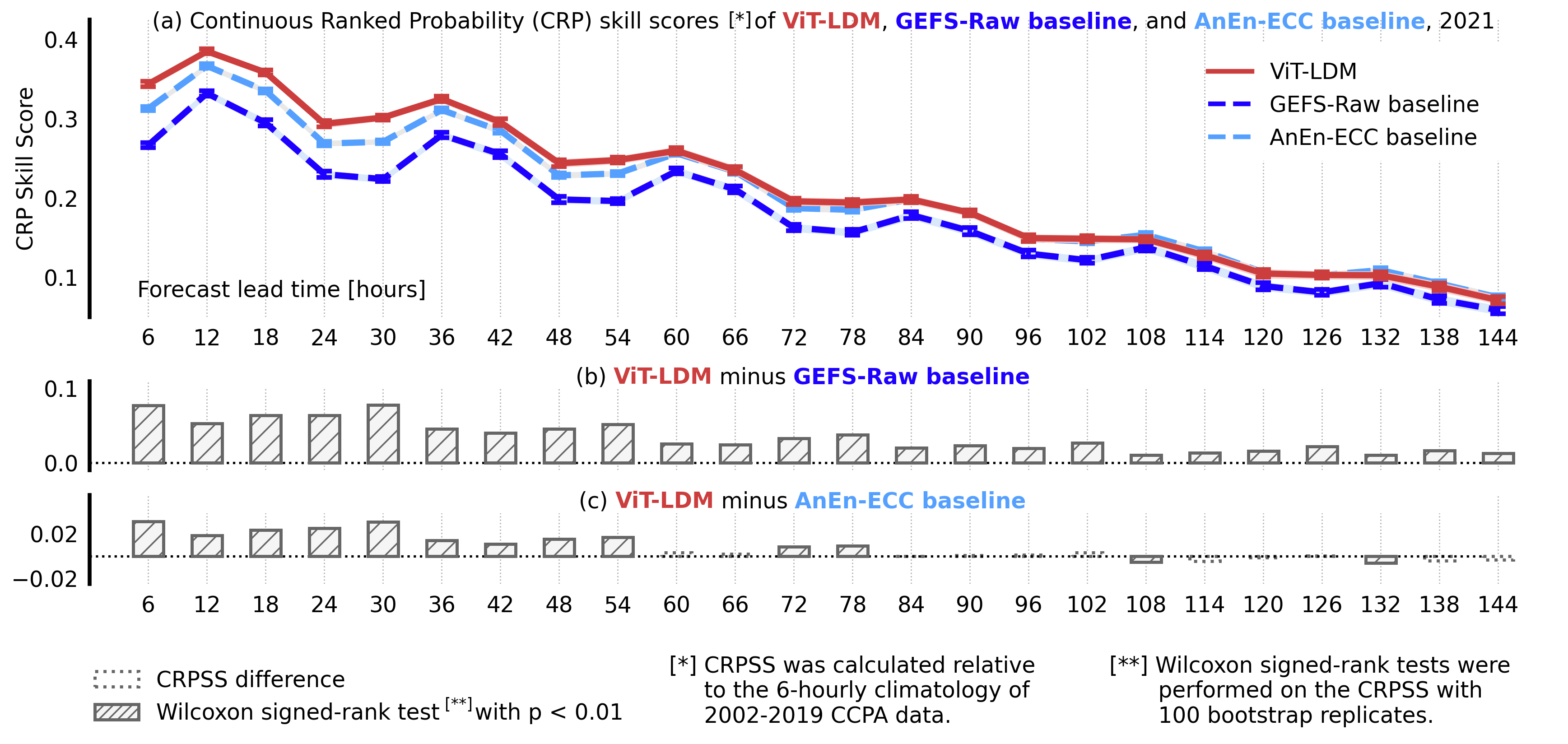}\\
\caption{Verification of ViT-LDM (red solid line), GEFS-Raw (blue dashed line), and AnEn-ECC (cyan dashed line) with Continuous Ranked Probability Skill Scores (CRPSSs; higher is better) in 2021. (a) Domain-wise averaged CRPSS curves by forecast lead times. (b) The CRPSS differences between ViT-LDM and GEFS-Raw. (c) The CRPSS differences between ViT-LDM and AnEn-ECC. CRPSS curves in (a) were averaged from 100 bootstrapped replicates with error bars representing the 95\% confidence intervals}\label{fig7}
\end{figure}

CRPSSs were averaged over all CONUS gridpoints and shown as functions of 6-hourly forecast lead times. CRPSS compares the entire predicted CDF, as represented by ensemble members, against the deterministic verification target. The CRPSS verification here is not focused on extreme precipitation events; rather, it measures the general forecast skill of the precipitation ensemble. 

ViT-LDM and AnEN-ECC performed better than the GEFS-Raw (Fig.~\ref{fig7}a), indicating that both post-processing methods can produce more skillful precipitation forecasts than the raw ensemble output. Their CRPSS gains were positive throughout but larger for shorter forecast lead times and smaller for 72-hour and longer lead times (Fig.~\ref{fig7}b). The reduced forecast skills in longer forecast lead times indicate that the limited predictability of GEFS APCP placed a strong impact on all post-processing methods. With the raw precipitation forecasts gradually diverging from the verification target on longer forecast lead times, it is difficult for post-processing methods to reconstruct the correct precipitation fields. The CRPSS differences between ViT-LDM and AnEn-ECC were statistically significant for the first 48 hours, with ViT-LDM performing better. For longer forecast lead times, the performance of AnEn-ECC was slightly superior to ViT-LDM (Fig.~\ref{fig7}c). This indicates that AnEn-ECC is a competitive baseline; it can produce statistically calibrated precipitation forecasts with improved CRPSS.

\begin{figure}[t]
\noindent\includegraphics[width=\textwidth]{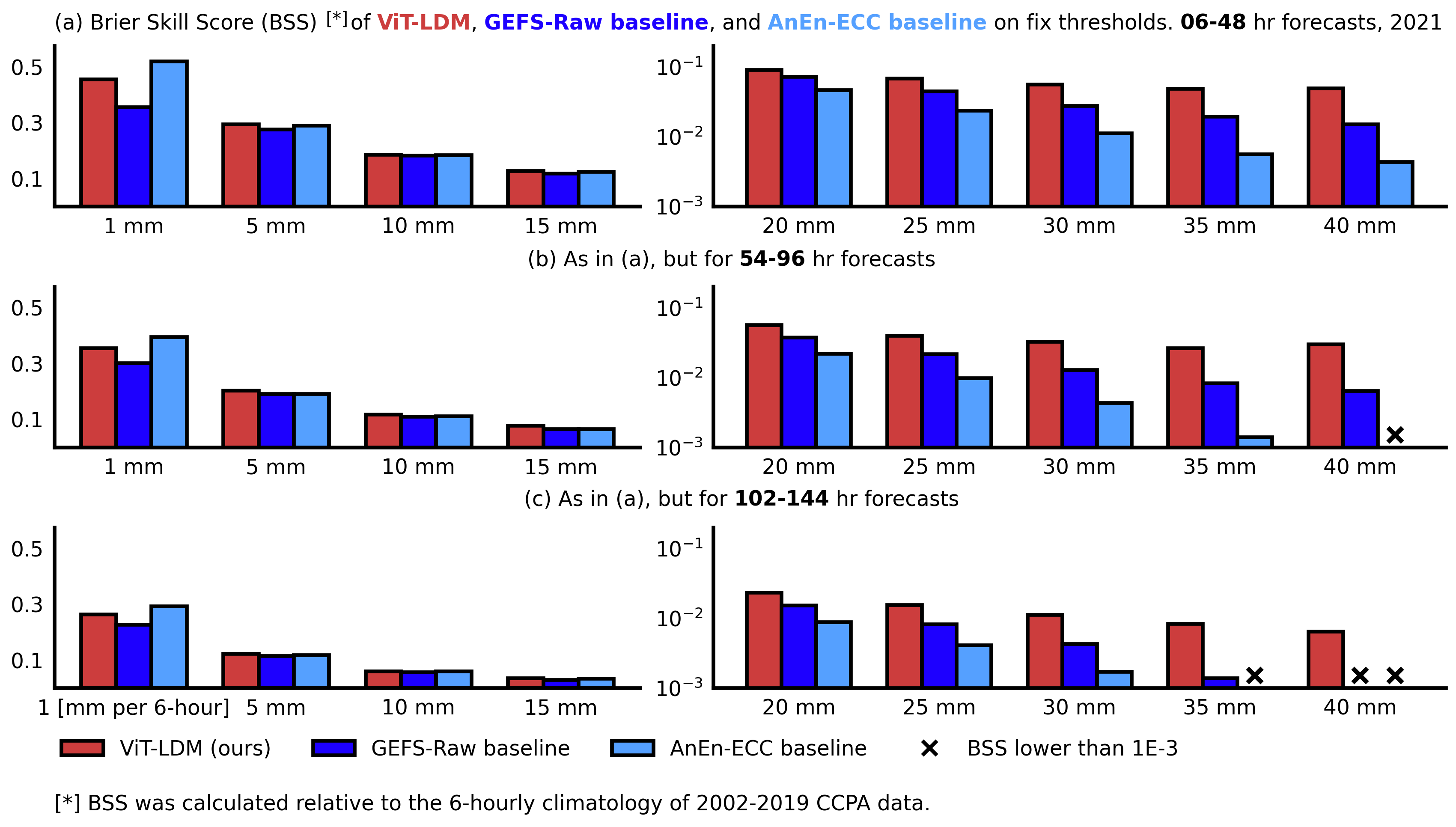}\\
\caption{Verifications of ViT-LDM (red bars), GEFS-Raw (blue bars), and AnEn-ECC (cyan bars) with Brier Skill Scores (BSS; higher is better) in 2021. (a) BSSs of precipitation events derived from 1 mm--40 mm per 6-hour thresholds and for 06-54-hour forecasts. (b) As in (a), but for 54-102-hour forecasts. (c) As in (a), but for 102-144-hour forecasts. ``x'' means the BSS is lower than 0.001.}\label{fig8}
\end{figure}

The BSSs of precipitation events computed from a set of fixed thresholds, ranging from 1 mm to 40 mm per 6 hours, were examined to provide further insights into the general performance of the precipitation ensembles. The lower end of these thresholds, such as 1 mm, 5 mm, and 10 mm per 6 hours, are related to mild and moderate precipitation events, whereas 20 mm per 6 hours and above characterizes heavy-to-extreme events. 

At lower thresholds, AnEn-ECC had the largest BSS among the three techniques (Fig.~\ref{fig8}). Its BSS at 1 mm per 6 hours events was $\approx$ 0.5, which also provided major contributions to the CRPSS increase in Fig.~\ref{fig7}a. For heavy-to-extreme events, however, the performance of AnEn-ECC decreased quickly with increasing threshold values, indicating that it is not an ideal option for post-processing extreme precipitation events. While the performance of ViT-LDM was suboptimal for precipitation events with lower thresholds, it was superior for heavy-to-extreme events, which increasing benefit as the threshold was increased. The good performance of ViT-LDM on extreme precipitation events will be examined further with reliability diagrams. Its suboptimal performance on mild and moderate precipitation events will also be discussed with explainability studies.

\subsection{Extreme precipitation verification with reliability diagrams}\label{sec43}

\begin{figure}[t]
\noindent\includegraphics[width=\textwidth]{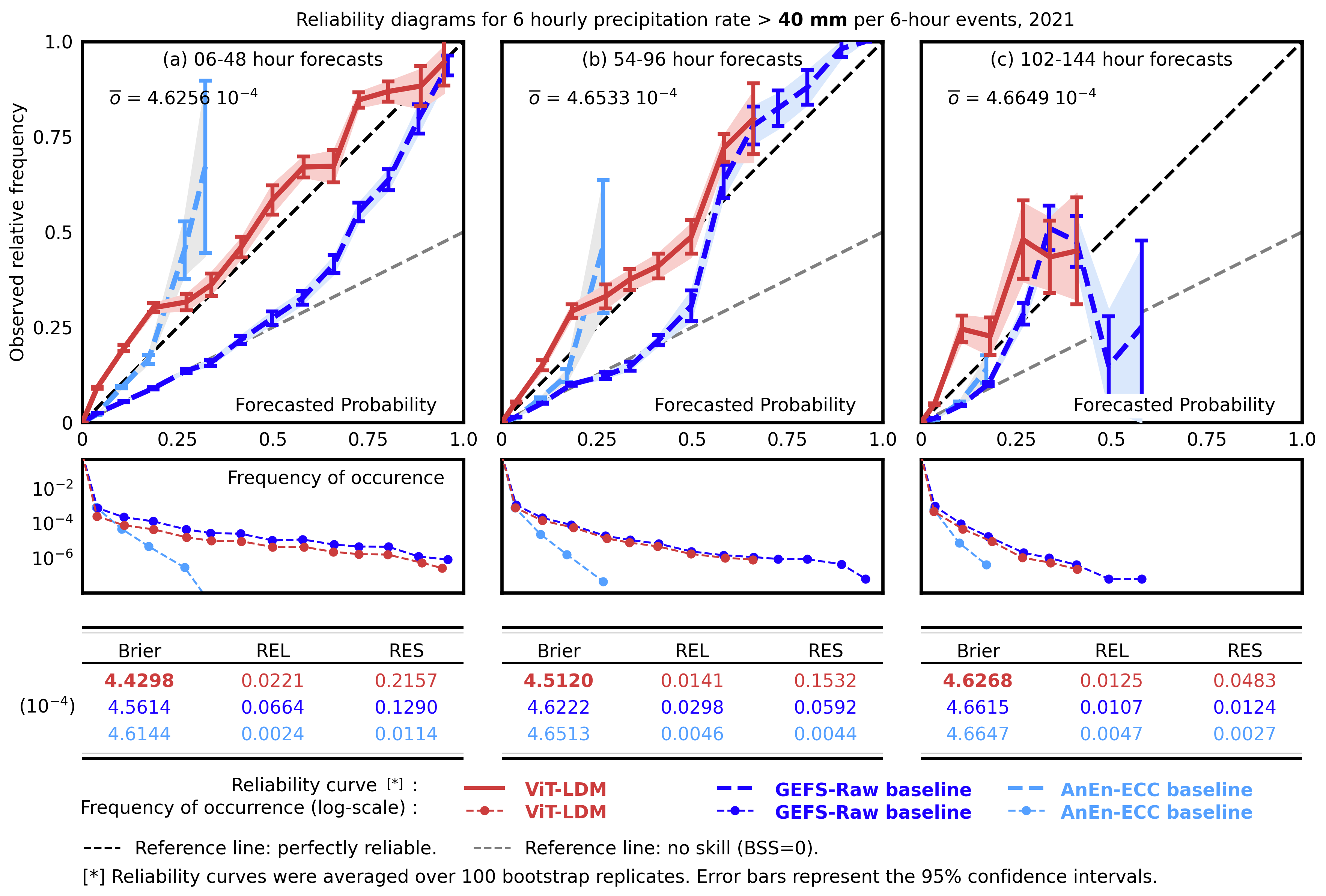}\\
\caption{Verification of forecasted 6-hourly extreme precipitation events with reliability diagrams (top), frequency of occurrence  (middle), and Brier score (``Brier''; lower is better) decompositions [bottom; reliability (``REL''; lower is better), resolution (``RES''; higher is better), and climatological uncertainty ($\overline{o}$)] in 2021. All scores were computed based on events of 6 hourly precipitation rates $>$ 40 mm per 6-hour threshold. In (a-c), metrics were averaged over 06-54, 54-102, and 102-144-hour forecasts, respectively. Dashed no-skill reference lines and perfect reliability diagonal reference lines are included. Calibration curves were averaged from 100 bootstrap replicates with error bars and color shades representing the 95\% confidence intervals.}\label{fig9}
\end{figure}

Reliability diagrams in Fig.~\ref{fig9} provided detailed calibration performance of all methods on forecasting 6 hourly extreme precipitation events, defined based on the fixed 40 mm per 6 hours threshold. As introduced in Section \ref{sec2}.\ref{sec21} and Fig.~\ref{fig1}, this threshold emphasizes forecast performance across the Great Plains and the Southeastern U.S., where extreme precipitation events are typically triggered by supercell thunderstorms, mesoscale convective systems, and other forms of small-scale convection. Producing both well-calibrated and sharp probabilistic precipitation predictions at this high of a threshold is a challenge at which most post-processing methods have struggled. 

The performance of AnEn-ECC on extreme precipitation events was found suboptimal in Fig.~\ref{fig8}. From the BS decompositions, it was revealed that AnEn-ECC improved the reliability from GEFS-Raw, but its resolution was too low, due to probabilities that rarely exceeded 30\%. In addition, the AnEn-ECC results were also underconfident, likely due to the fact that 40 mm per 6 hours is roughly the 99.6th percentile of the verified area (Fig.~\ref{fig1}e and f). These events may not be represented well within the GEFS reforecast training set. 

The GEFS-Raw forecasts exhibited better resolution than the AnEn-ECC forecasts but the GEFS-Raw forecasts were also unreliable; its calibration curve stayed close to the ``no-skill'' reference line. The GEFS-Raw has improved resolution compared to AnEn-ECC because it over-predicted extreme precipitation events among all its members, which resulted in the probabilistic forecasts being distinguishable from the climatological mean. However, the GEFS-Raw forecasts were found to have poor reliability as their forecasted probabilities often did not co-occur with observed extreme precipitation events.

The ViT-LDM forecasts showed the best performance in this verification. Its 06-54-hour calibration performance was impressive, with the number of high-probability forecasts comparable to that of the GEFS-Raw, and a reliability curve followed the perfectly reliable line. For longer forecast lead times, particularly 102-144 hours, the resolution of ViT-LDM decreased, mainly due to the reduced predictability of these extreme events. Nonetheless, ViT-LDM still outperformed the two baselines.

Overall, for extreme precipitation events closely related to deep and intense convection in the Great Plains and the Southeastern US, ViT-LDM exhibited excellent calibration performance for short forecast lead times and clearly outperformed the two baselines for all verified forecast lead times. This result is also aligned with \citet{sha2024generative}, which revealed the good performance of generative AI in predicting severe weather events in this area. 

\begin{figure}[t]
\noindent\includegraphics[width=\textwidth]{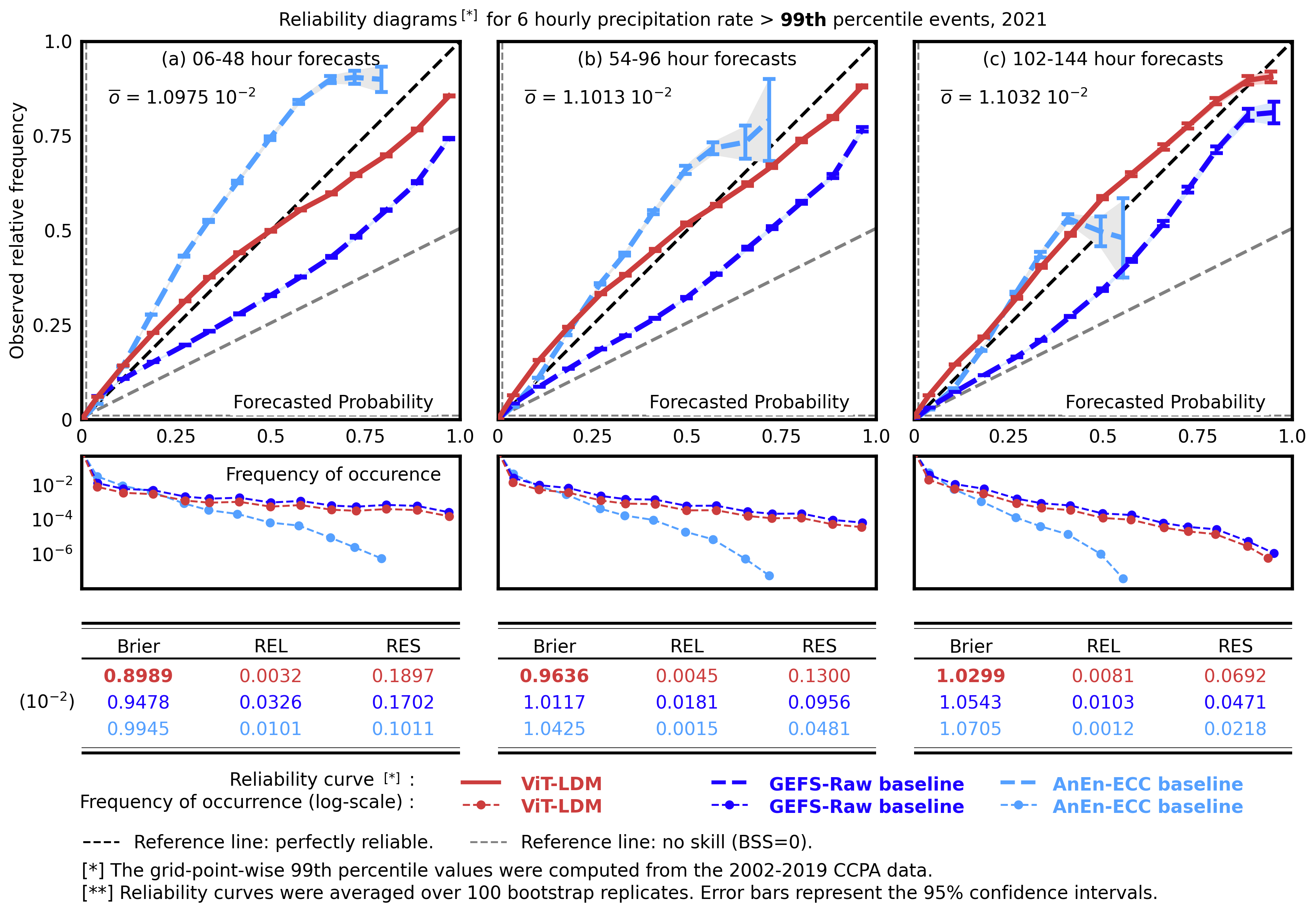}\\
\caption{As in Fig.~\ref{fig9}, but for extreme events of 6-hourly precipitation rates $>$ gridpoint-wise 99th percentile thresholds. Note that $\overline{o}$ is not strictly equal to 0.01 because it was derived from the 2000-2019 CCPA climatology, not from the 2021 verification period.}\label{fig10}
\end{figure}

Fig.~\ref{fig10} examines forecast reliability for extreme precipitation events identified based on the gridpoint-wise 99th percentile thresholds. Similar to the 40-mm-based verification in Fig.~\ref{fig9}, the resolution of AnEn-ECC was too low, which reduced its calibration performance. Since the actual precipitation rate of 99th percentile thresholds were typically lower than 40 mm per 6 hours, AnEn-ECC generated more large probabilities (Fig.~\ref{fig9}). The GEFS-Raw was capable of issuing higher probabilities for extreme precipitation events as well, however, probabilities were often overforecast and its reliability curves stayed around the ``no-skill'' reference line.

Among the three methods, the ViT-LDM had the best calibration. Its reliability was comparable to the AnEn-ECC baseline but preserved the resolution of the GEFS-Raw forecasts. The latter can be further confirmed by the frequency of occurrence plots, where the number of high-probability extreme precipitation forecasts issued by the ViT-LDM was comparable to that of the GEFS-Raw. Meanwhile, the AnEn-ECC rarely produced extreme precipitation probabilities $>$ 0.5. Overall, for 6 hourly extreme precipitation events defined based on 99th percentile thresholds, the post-processed ensemble trajectories produced by ViT-LDM were verified to be skillful compared to the two baseline forecasts.

\begin{figure}[t]
 \noindent\includegraphics[width=\textwidth]{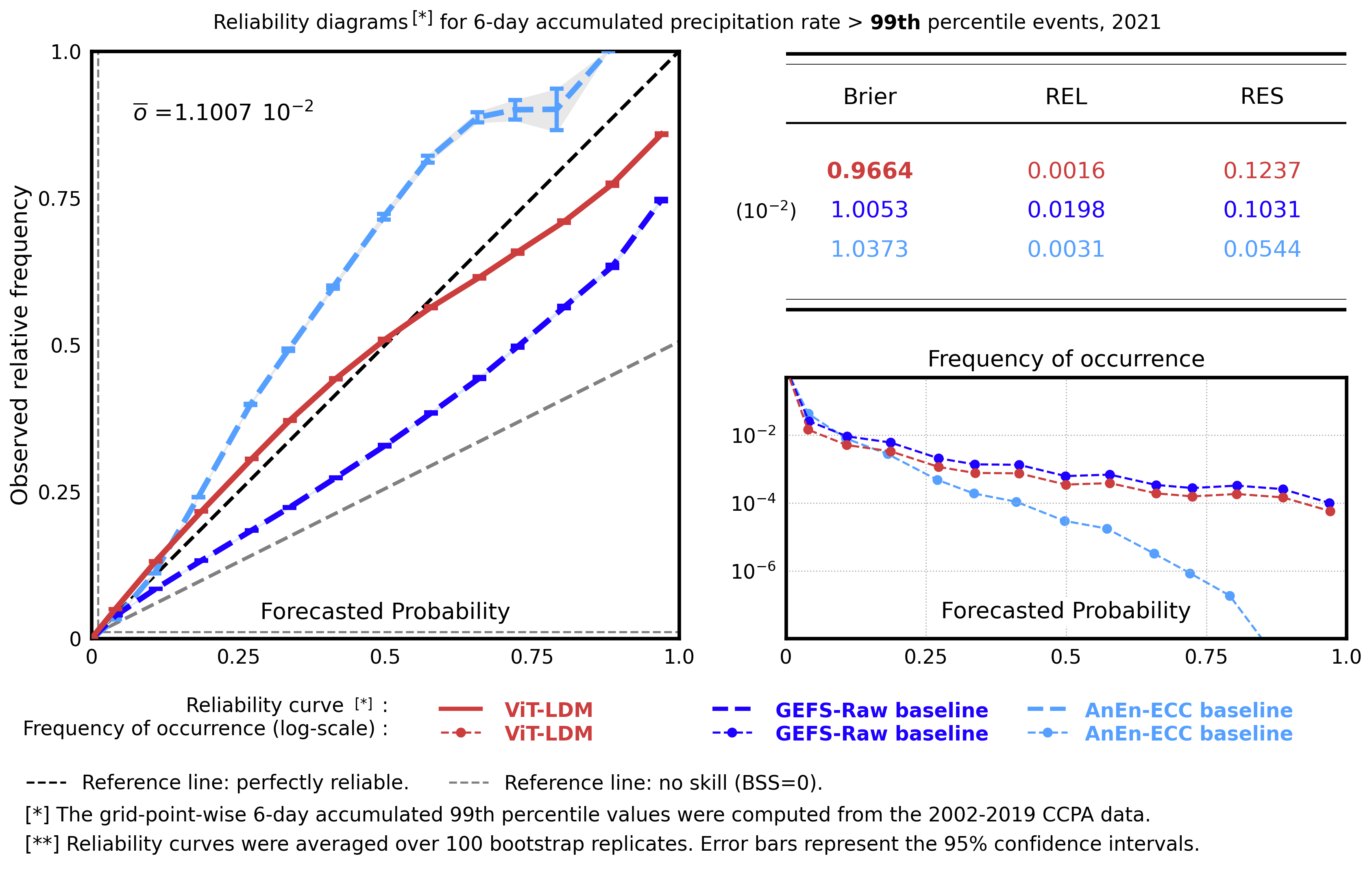}\\
 \caption{As in Fig.~\ref{fig9}, but for extreme events of 6-day accumulated precipitation amount $>$ gridpoint-wise 99th percentile thresholds.}\label{fig11}
\end{figure}

Finally, we examine the reliability of 6-day accumulated precipitation greater than the gridpoint-wise 99th percentile (Fig.~\ref{fig11}). Temporally aggregated precipitation forecasts are sensitive to the spatiotemporal co-variability of the forecasted trajectories. Thus, this verification examines how well these methods can produce spatiotemporally consistent forecasts. In addition, it is also a good indicator of the usefulness of post-processing methods in real-world scenarios where end-users can be warned of a sequence of incoming extreme precipitation events.

Based on the shape and position of reliability curves in Fig.~\ref{fig11}, all methods were as reliable as they were for 6 hourly forecast lead times, indicating that ViT-LDM and the two baselines produced spatiotemporally consistent forecast trajectories. The reliability and resolution of the 6-day accumulated forecasts were consistent with the performance of the short lead-time 6--54-hour forecasts. This is because the timing error of extreme events was largely eliminated when the entire trajectory was aggregated into a single time frame \citep[e.g.][]{jeworrek2021wrf}. 

For the two baselines, their spatiotemporal consistency was expected because ECC re-assembles AnEn members by using the GEFS-Raw as dependence templates \citep{schefzik2013uncertainty}, and the GEFS-Raw, as produced for a physics-based numerical model, is spatiotemporal consistent. For ViT-LDM, its spatiotemporal consistency was confirmed in this verification, and it outperformed the two baselines with the best reliability and resolution decompositions. This result shows the effectiveness of 3-D ViT and LDM on bias-correcting and generating forecast trajectories that characterizes the evolution of extreme precipitation events well, and these trajectories are practical to be used as 6-day guidance.

%Evaluations of uncertainty quantification
\subsection{VQ-VAE latent space visualization and explainability studies}\label{sec44}

\begin{figure}[t]
\noindent\includegraphics[width=\textwidth]{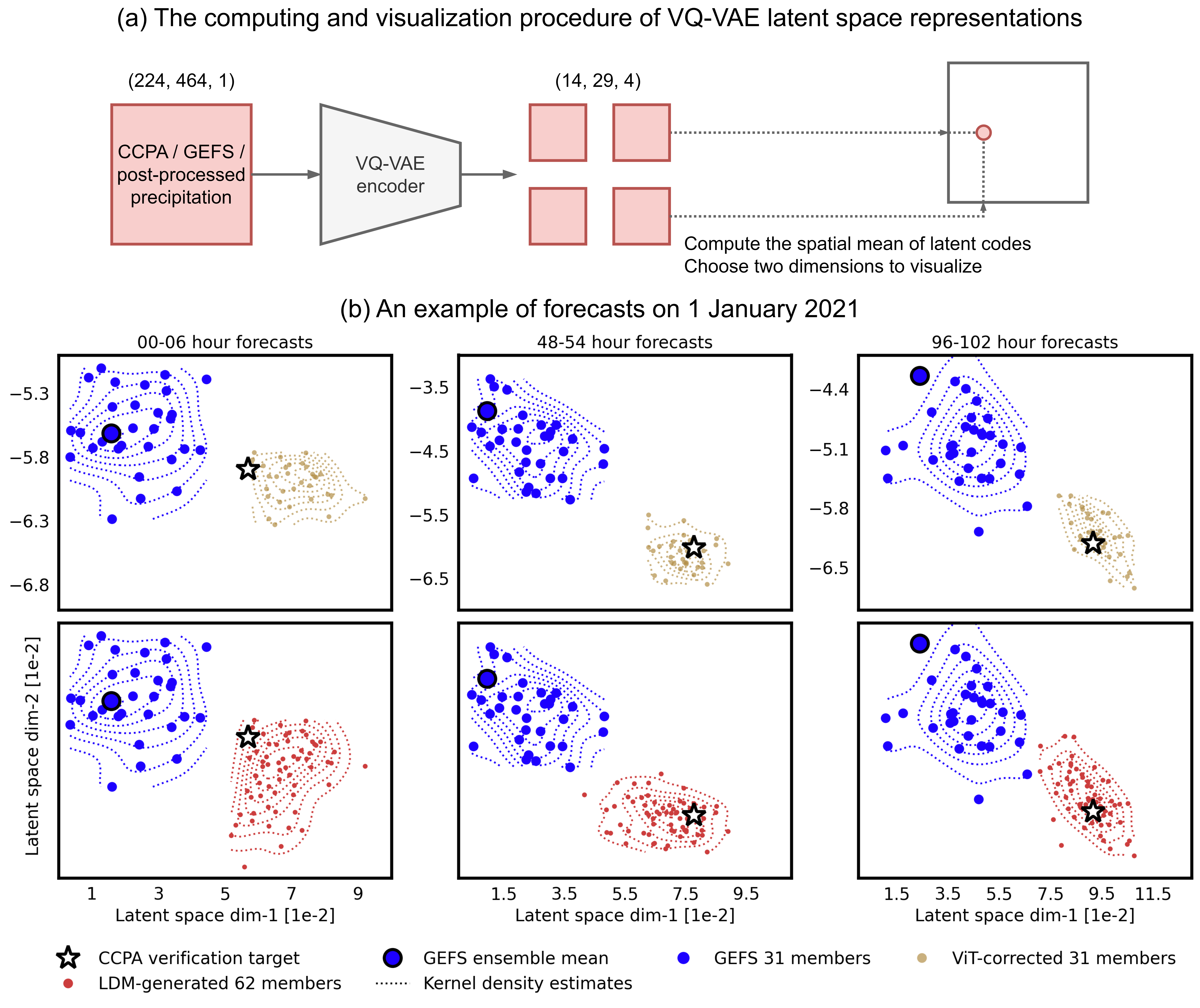}\\
\caption{(a) The schematics of the VQ-VAE latent space visualization. (b) Visualization examples on 1 January 2021 with (a) 00-06, (b) 48-54, and (c) 96-102-hour forecasts, respectively. Small and large blue dots are the latent space representations of the original GEFS ensemble members and the ensemble mean, respectively. Yellow and red dots are the representations of 3-D ViT outputs and ViT-LDM outputs. Star symbols are the representations of the CCPA verification target. Dashed lines were produced from kernel density estimates.}\label{fig12}
\end{figure}

In this section, the predictive behavior of the 3-D ViT and LDM were examined within the VQ-VAE latent space. The purpose of this study is to identify the contribution of the two neural networks in precipitation post-processing and ensure that their performances were attributed to decisions rather than overfitting or artifacts.

The same 1 January 2021 extreme precipitation events as in Fig.~\ref{fig6}, but with 00-06, 48-54, and 96-102 hour forecasts were selected, and the technical steps of their explainability studies were introduced in Fig.~\ref{fig12}a: The encoder projects the raw GEFS forecasts, CCPA targets, and the outputs of the two neural networks into the VQ-VAE latent space. Two latent space dimensions out of four were selected, and their averaged codebook values were computed and visualized on 2-D axes. 

In Fig.~\ref{fig11}b, The latent space representations of the GEFS APCP forecasts and the CCPA targets exhibited large spatial differences (cf. blue dots and star symbols in Fig.~\ref{fig12}b). Such differences may not be identifiable in the real space with 224-by-464 gridpoints, however, when projected to a condensed and regularized VQ-VAE latent space, different positions were assigned for forecasts and analysis fields. That said, with the disentanglement property of a pre-trained VQ-VAE, GEFS APCP forecasts and the CCPA targets are clearly separable within the latent space. 

The separations of forecasts and analysis further revealed the decision-making process of the 3-D ViT; it relocates each GEFS APCP member from a forecast-oriented representation to an analysis-oriented representation (cf. blue dots and yellow dots in Fig.~\ref{fig12}b). When all the GEFS members were post-processed in this way, they would stay around the position of the CCPA target; therefore, the overall CRPSS would expect an increase. Section \ref{sec3}.\ref{sec32} mentioned the training procedure of the 3-D ViT; it was trained using the 06-54-hour GEFS ensemble mean but applied to individual members and all forecast lead times. This training strategy can be explained in Fig.~\ref{fig12}b. For short forecast lead times, the latent space representation of the GEFS ensemble mean was surrounded by all its ensemble members well, which means the learned relationships between the ensemble mean and the CCPA target can be applied to individual members directly. For longer forecast lead times, the latent space position of the GEFS ensemble mean no longer stayed close to its ensemble members (cf. large and small blue dots in Fig.~\ref{fig11}b for different forecast lead times), so it has lost its ability on representing the bias correction relationships between individual ensemble members and the CCPA target. In addition, the relative positions of the GEFS forecasts and the CCPA targets stayed roughly the same for all visualized forecast lead times, which means the learned bias correction relationships for short forecast lead times can potentially be generalized to longer lead times. Thus, training 3-D ViT on short forecast lead times and apply to a longer range of hours is a valid option.

The role of the diffusion model and the limitation of ViT-based post-processing can be identified from this explainability study. 3-D ViT is a deterministic neural network; it relocates GEFS ensemble members within the VQ-VAE latent space to achieve bias correction, but the relocated members stayed very close to each other, and the spatial coverage of their representations was smaller than the raw GEFS members. We suspect these closely clustered members may have amplified the over-prediction of the GEFS on drizzle forecasts and caused the suboptimal performance on calibrating 1 mm and 5 mm per 6-hour events. The diffusion model showed the ability to enlarge such spatial coverage. Its generative ensemble preserved the latent space locations of the 3-D ViT outputs while expanding their spatial coverage (cf. yellow and red dots in Fig.~\ref{fig12}b).). The spatial expansion is expected to improve the overall CRPSS performance because the CRPS computation rewards inter-member differences when the mean absolute error is preserved \citep{grimit2006continuous}. In addition, the spatial expansion also generated a few outliers from the 3-D ViT outputs. As discussed in \citet{li2024generative}, generated outliers are connected to possible scenarios for the evolution of extreme weather, thus benefiting the calibration of extreme precipitation events.

\begin{table}[t]
\caption{Ablation studies of 3-D ViT only and the full ViT-LDM predictions, contrasted by the GEFS-Raw baseline in 2021. Domain-wise CRPS (lower is better) and BS of 6-hourly precipitation rate $>$ gridpoint-wise 99th percentile events (lower is better) were applied as metrics.}\label{tab1}
\centering
\begin{tabularx}{\textwidth}{CCCC}
\hline\hline
\multicolumn{4}{c}{Domain-wise CRPS, 2021} \\
\hline
 & 06-48 hours & 54-96 hours & 102-144 hours \\
\hline
GEFS-Raw & 0.363 & 0.404 & 0.444 \\
ViT only & 0.358 & 0.398 & 0.442 \\
ViT-LDM  & 0.334 & 0.389 & 0.436 \\
\hline
\hline
\multicolumn{4}{c}{BS of 6-hourly precipitation rate $>$ 99th percentile events, 2021 ($10^{-2}$)} \\
\hline
 & 06-48 hours & 54-96 hours & 102-144 hours \\
\hline
GEFS-Raw & 0.948 & 1.012 & 1.054 \\
ViT only & 0.917 & 0.981 & 1.042 \\
ViT-LDM  & 0.899 & 0.964 & 1.030 \\
\hline
\end{tabularx}
\end{table}

The predictive behaviors of ViT-LDM were further examined by removing certain components  Based on the verification set performances in Table \ref{tab1}, both the two neural networks contributed to the ensemble post-processing, and their contributions were comparable for the estimation of extreme precipitation events. For the general performance as measured by CRPS, the contribution of LDM is relatively larger than that of the 3-D ViT. This is aligned with the explainability studies, where the generative members were found to represent larger areas within the VQ-VAE latent space.

\begin{figure}[t]
 \noindent\includegraphics[width=\textwidth]{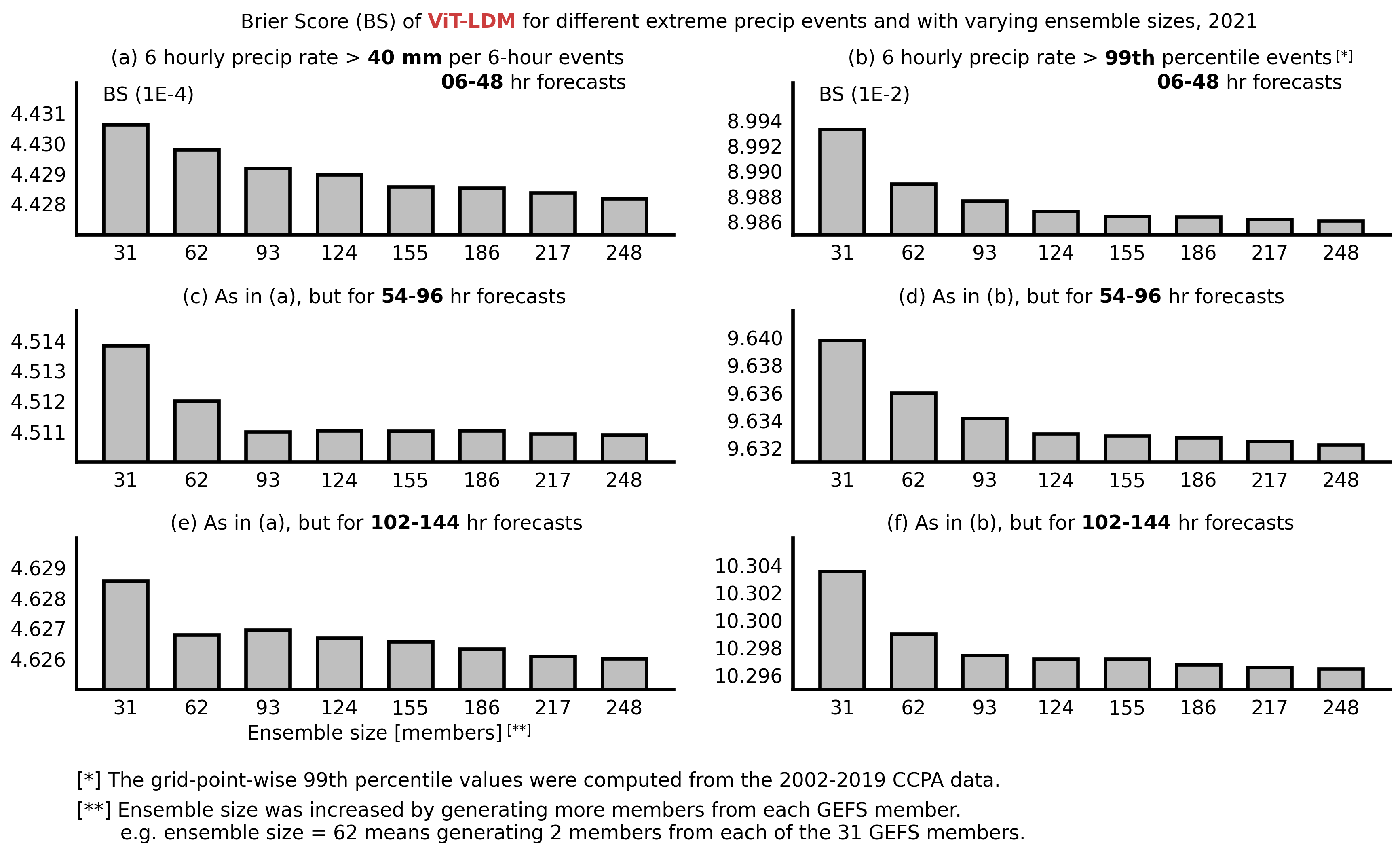}\\
 \caption{The Brier Score (BS; lower is better) performance of ViT-LDM on 6 hourly extreme precipitation events and with varying ensemble sizes. The ensemble sizes were increased by generating more members from each GEFS-Raw member.}\label{fig13}
\end{figure}

The verification and explainability studies confirmed the effectiveness of ViT-LDM as a post-processing framework that supports the probabilistic estimation of extreme precipitation events. Here, the potential benefits of ViT-LDM are examined by comparing its performance across different extreme precipitation events and ensemble sizes. As shown in Fig.~\ref{fig13}, generating more ensemble members would typically lead to more skillful probabilistic forecasts of extreme precipitation events. For short forecast lead times and more extreme thresholds (i.e., 40 mm per 6-hour), these benefits are more pronounced, with the ensemble size increase and the BS decrease exhibiting a close-to-linear relationship (Fig.~\ref{fig13}a and b). For longer forecast lead times and less extreme thresholds (i.e., 99th percentile), diminishing marginal effects were observed. When the ensemble size exceeds 100 members, the benefit of adding more members becomes limited (Fig. 13c-f).

Two reasons may explain such phenomena: (1) For less extreme thresholds, 100 members may have provided a sufficiently large ensemble space for calibration; (2) for 40 mm per 6-hour events in longer forecast lead times, the forecasts of GEFS-Raw and ViT-LDM may lack sufficient predictability  (c.f. Fig.~\ref{fig13}a and c), so generating more ensemble members of the same quality does not improve performance. In summary, the potential benefits of ViT-LDM is considered the largest for short-term extreme precipitation events with higher/stricter thresholds. This finding can be considered for the research and operation of generative-AI-based methods in the future.

% Discussion
\section{Discussion}\label{sec5}

This research utilized generative AI for the ensemble post-processing of extreme precipitation events. Two research questions were examined. The first was related to the feasibility of generative AI and its performance in forecasting extreme precipitation. The implementation of generative AI in this study was successful, and its technical approach was similar to that of \citet{li2024generative} and \citet{zhong2023fuxi}, which applied diffusion models using forecasted fields as conditional inputs. An important technical choice of ViT-LDM that differs from other research is the use of latent space projection, which reduced the overall data sizes and simplified the training of 3-D ViT and LDM. In the early stage of this research, the authors experimented with 3-D ViTs configured with larger patch sizes and without using a VQ-VAE. Strong checkerboard artifacts were identified with this choice. The VQ-VAE-based latent space projection and patch size 1 were then proposed, leading to the success of this generative AI application. Elevation and climatology inputs were incorporated into the ViT-LDM pipeline as background information. This option benefits the generation of precipitation fields with better quality, and its effectiveness has been discussed in other downscaling \citep[e.g.][]{sha2020deepa,sha2020deepb,wang2021deep} and bias-correction\citep[e.g.][]{sha2022hybrid} studies. The performance of ViT-LDM in predicting extreme precipitation events were verified to be better than the AnEn-ECC baseline, a set of widely used post-processing techniques, as well as the original GEFS ensemble. The good performance of ViT-LDM is especially evident from the verification of 40 mm per 6-hour events, which many existing post-processing methods cannot calibrate properly. We think generative AI has the potential to be applied to the ensemble generation of regional precipitation forecasts and improve the prediction of extreme precipitation events.

The second research question focused on the explainability of ViT-LDM. Here, the example-based explainability studies were conducted. An important finding of this study is that the VQ-VAE latent space is capable of separating GEFS forecasts and the CCPA targets, which further explains the effectiveness of 3-D ViT in relocating the forecast-oriented representations. The role of LDM was identified as expanding the spatial area of latent space representations, which improved both the general post-processing performance and the calibration of extreme precipitation events. The explainability of AI-based forecast post-processing methods are generally lacking. For many neural-network-based post-processing methods, it is unclear how they have improved the quality of numerical forecasts. This research provided an example of exploring the decision-making mechanism of these neural networks. In addition, explorations of performance with varying ensemble sizes in Section \ref{sec4}.\ref{sec44} have also brought evidence and new insights on the potential benefits of implementing generative AI in extreme weather problems.

The ViT-LDM was found to be suboptimal for calibrating mild and moderate precipitation events, defined by 1 mm and 5 mm per 6-hour thresholds. Future work could be conducted to tackle this challenge. Based on the explainability results, we think the order of the two post-processing steps can be changed. That is, generative AI can be applied to the raw forecast directly, and the resulting generative ensemble can be re-calibrated by another post-processing method. This choice avoids the use of a deterministic post-processing neural network and enables the flexibility of implementing ensemble-based calibration methods. 

% Conclusions
\section{Conclusions}\label{sec6}

A novel post-processing method, ViT-LDM, was proposed by incorporating a Vector Quantised-Variational AutoEncoder (VQ-VAE), a 3-D Vision Transformer (ViT), and a Latent Diffusion Model (LDM). The method takes 6-hourly precipitation forecasts from numerical ensembles as inputs and generates post-processed trajectories that are skillful for the probabilistic estimation of extreme precipitation events. The 3-D ViT aims to reduce conditional bias from the original numerical ensemble, while the LDM produces an expanded generative ensemble that better characterizes extreme events.

The method was trained using the Global Ensemble Forecast System version 12 (GEFSv12) reforecasts as inputs and the Climate-Calibrated Precipitation Analysis (CCPA) as targets from 2002 to 2019 and tested with the operational GEFSv12 over the Conterminous United States (CONUS) from 1 January 2021 to 31 December 2021. Verification results showed that the method generated skillful precipitation forecast trajectories, as indicated by Continuous Ranked Probabilistic Skill Scores (CRPSSs) and Brier Skill Scores (BSSs).  Its calibration performance for extreme precipitation events was superior to that of the operational GEFS and the combination of Analog Ensemble (AnEn) and Ensemble Copula Coupling (ECC). Reliability diagrams demonstrated that the probabilistic extreme precipitation forecasts of ViT-LDM were as reliable as that of the AnEn-based calibrations but with better resolution scores. For the verification of 6-day accumulated extreme precipitation events, ViT-LDM maintained the same good reliability and resolution seen across the individual 6-hourly forecast lead times, indicating that its generated trajectories were spatiotemporally consistent and could be aggregated to provide multi-day forecast guidance.

Explainability studies were conducted to examine the decision-making process of ViT-LDM and provided evidence on the potential benefits of implementing generative methods in severe weather problems. These studies revealed that the VQ-VAE latent space provided good separations between the GEFS forecasts and the CCPA analysis, while the 3-D ViT was capable of relocating the latent space representations of the raw GEFS members to achieve bias correction. It was also confirmed that the latent space representations of the LDM-generated members were clustered around the location of the CCPA verification target with an enlarged spatial coverage. This enlarged cluster with expanded ensemble size improved the characterization of extreme precipitation events. A potential weakness of ViT-LDM was its suboptimal calibration performance for mild and moderate precipitation events, attributed to the smoothness effect of the VQ-VAE decoder and the deterministic nature of the 3-D ViT bias-correction network. Possible solutions were discussed as future research directions.

In summary, ViT-LDM leverages a generative Artificial Intelligence (AI) approach for extreme precipitation forecasts.  It produces skillful and spatiotemporally consistent precipitation forecast trajectories, bridging the gap between limited numerical ensemble sizes and the need for large ensemble sets to assess extreme precipitation events. More broadly, it provides a framework for implementing generative AI methods to address weather forecasting challenges.

\clearpage
%%%%%%%%%%%%%%%%%%%%%%%%%%%%%%%%%%%%%%%%%%%%%%%%%%%%%%%%%%%%%%%%%%%%%
% ACKNOWLEDGMENTS
%%%%%%%%%%%%%%%%%%%%%%%%%%%%%%%%%%%%%%%%%%%%%%%%%%%%%%%%%%%%%%%%%%%%%
\acknowledgments
The authors thank Dr. Yan Luo, I.M. Systems Group, Inc. (IMSG), NOAA, for the archived CCPA dataset. This material is based upon work supported by the National Center for Atmospheric Research (NCAR), which is a major facility sponsored by the National Science Foundation (NSF) under Cooperative Agreement No. 1852977. This research was supported by NOAA OAR grant NA19OAR4590128, the NSF NCAR Short-term Explicit Prediction Program, and NSF Grant No. ICER-2019758. Supercomputing support was provided by NSF NCAR Cheyenne and Casper (Computational and Information Systems Laboratory, CISL 2020). The authors also thank Dr. John Schreck at NSF NCAR for his feedback.
%  Keep acknowledgments (note correct spelling: no ``e'' between the ``g'' and
% ``m'') as brief as possible. In general, acknowledge only direct help in
%  writing or research. Financial support (e.g., grant numbers) for the work done, 
%  for an author, or for the laboratory where the work was performed must be 
%  acknowledged here rather than as footnotes to the title or to an author's name.
%  Contribution numbers (if the work has been published by the author's institution 
%  or organization) should be placed in the acknowledgments rather than as 
%  footnotes to the title or to an author's name.

%%%%%%%%%%%%%%%%%%%%%%%%%%%%%%%%%%%%%%%%%%%%%%%%%%%%%%%%%%%%%%%%%%%%%
% DATA AVAILABILITY STATEMENT
%%%%%%%%%%%%%%%%%%%%%%%%%%%%%%%%%%%%%%%%%%%%%%%%%%%%%%%%%%%%%%%%%%%%%
% 
%
\datastatement
The data pre-processing, neural network training, and data visualization code of this research can be found at: \url{https://github.com/yingkaisha/ViT_Diffusion_GEFS}. The GEFSv12 forecasts and reforecasts are available at \url{https://aws.amazon.com/marketplace/pp/prodview-qumzmkzc2acri} and \url{https://registry.opendata.aws/noaa-gefs-reforecast/}, respectively. The long-term CCPA data of this research is archived in the NOAA supercomputing system; readers may contact Dr. Jun Du at the Environmental Modeling Center (EMC), NOAA, for details. The one-week near-real-time CCPA data is available at: \url{https://ftp.ncep.noaa.gov/data/nccf/com/ccpa/prod/}.

%  The data availability statement is where authors should describe how the data underlying 
%  the findings within the article can be accessed and reused. Authors should attempt to 
%  provide unrestricted access to all data and materials underlying reported findings. 
%  If data access is restricted, authors must mention this in the statement. See
%  {http://www.ametsoc.org/PubsDataPolicy} for more info.

%%%%%%%%%%%%%%%%%%%%%%%%%%%%%%%%%%%%%%%%%%%%%%%%%%%%%%%%%%%%%%%%%%%%%
% APPENDIXES
%%%%%%%%%%%%%%%%%%%%%%%%%%%%%%%%%%%%%%%%%%%%%%%%%%%%%%%%%%%%%%%%%%%%%
%
%% If only one appendix, use

\appendix

\appendixtitle{Improving Ensemble Extreme Precipitation Forecasts using Generative Artificial Intelligence: supplemental material}

The main article introduced ViT-LDM, a post-processing framework that incorporates a Vector-Quantized Variational Autoencoder (VQ-VAE), a 3-D Vision Transformer (ViT), and a Latent Diffusion Model (LDM) for the ensemble generation of the Global Ensemble Forecast System (GEFS) total precipitation forecasts (APCP). This supplemental document includes the hyperparameter choices and training procedures of the above neural networks. Descriptions of the additional information here are paired to Section 3.a-c of the main article.

\subsection{VQ-VAE}

The VQ-VAE contains two parts: an encoder-decoder as its main structure and an output section that takes elevation and precipitation climatology as inputs to produce refined precipitation fields. The two parts were trained separately. Table \ref{tab1_a} summarizes the hyperparameters that were considered for the main structure. These hyperparameters were examined during the VQ-VAE training stage and selected based on the validation set performance after a fixed number of training epochs; similar validation-set-based hyperparameter selections were applied to other models in this research. VQ-VAE may exhibit convergence problems. A few solutions were proposed during its training stage: (1) pre-train a vanilla autoencoder to initialize the VQ-VAE weights. (2) Freeze the VQ-VAE decoder and fine-tune the encoder and VQ-layer. (3) Cosine annealing with warm re-start \citep{loshchilov2016sgdr}.

\begin{table}[h]
\caption{Identified hyperparameters of the VQ-VAE and their initial search options.}\label{tab1_a}
\vspace{1ex}
\centering
\begin{tabularx}{\textwidth}{llC}
\\[-1.8ex]\hline 
\hline \\[-1.8ex]
Name & Hyperparameter options & Choice \\
\hline
\multirow{2}*{Down- and upsampling rate} & (1) 2-by-2 with 4 layers & \multirow{2}*{(2)} \\
 & (2) 4-by-4 with 2 layers & \\

\multirow{3}*{Hidden layer dimensions} & (1) 2 layers: [64, 128] & \multirow{3}*{(1)} \\
 & (2) 2 layers: [96, 96] & \\
 & (3) 4 layers: [64, 96, 128, 160] & \\
 
\multirow{2}*{Normalization options} & (1) Batch normalization & \multirow{2}*{(1)} \\
 & (2) Layer normalization & \\

\multirow{2}*{Upsampling operation} & (1) Transpose convolution & \multirow{2}*{(2)} \\
 & (2) Linear interpolation with convolution & \\

\multirow{2}*{Activation function} & (1) ReLU & \multirow{2}*{(2)} \\
 & (2) GELU & \\

\multirow{2}*{Learning rate schedule} & (1) 1E-4 with cosine annealing & \multirow{2}*{(2)} \\
 & (2) Fixed 1E-5 & \\
\hline
\end{tabularx}
\end{table}

\noindent
The fraction of the commitment loss was set as $\beta=0.25$. The separated output section of VQ-VAE was trained with the frozen main structure. Its hyperparameters were considered based on Table \ref{tab1} directly. Adam optimizers \citep{kingma2014adam} were used throughout. After training, the VQ-VAE may occasionally produce negative precipitation values. As introduced in Section 3.a, a 0.1 mm truncation was applied to solve the problem.

\subsection{3-D ViT}

The 3-D ViT was trained using GEFS reforecast APCP ensemble mean as inputs and the CCPA data as targets. It was validated using 5 GEFS reforecast members on randomly selected initializations and Continuous Ranked Probability Score (CRPS) as the metric. The training process relies on the pre-trained VQ-VAE. Tabel \ref{tab2} summarizes the hyperparameters of 3-D ViT.

\begin{table}[h]
\caption{Identified ViT hyperparameters and their initial search options.}\label{tab2}
\vspace{1ex}
\centering
\begin{tabularx}{\textwidth}{llC}
\\[-1.8ex]\hline 
\hline \\[-1.8ex]
Name & Hyperparameter options & Choice \\
\hline
\multirow{2}*{Vision Transformer backbone} & (1) the original ViT & \multirow{2}*{(1)} \\
 & (2) SwinT version 2 & \\

\multirow{3}*{Patch size} & (1) (time=1, lat=1, lon=1) & \multirow{3}*{(1)} \\
 & (2) (time=2, lat=2, lon=2) & \\
 & (3) (time=4, lat=1, lon=1) & \\
 
\multirow{3}*{Embedded dimension} & (1) 64 & \multirow{3}*{(2)} \\
 & (2) 128 & \\
 & (3) 256 & \\
 
\multirow{2}*{Number of ViT blocks and attention heads} & (1) 4 blocks with 8 attention heads & \multirow{2}*{(2)} \\
 & (2) 8 blocks with 4 attention heads & \\

\multirow{2}*{Learning rate schedule} & (1) 1E-4 with cosine annealing & \multirow{2}*{(1)} \\
 & (2) Fixed 1E-5 & \\
\hline
\end{tabularx}
\end{table}

\noindent
For patch sizes, lat=1 and lon=1 are highly preferred because this avoids the checkerboard artifacts. We have experimented with larger spatial patch sizes followed by additional convolutional layers, and the results were suboptimal. We did not experiment with more attention head options because this feature is expensive to compute. Shift-window-based Transformers (SwinTs; \citealt{liu2021swin}) were tested during the hyperparameter search, but it did not improve the performance. A possible reason is that the embedded patches have both spatial and forecast lead time dimensions; conventional ViT with global-scale cross-attentions may work better on capturing the spatiotemporal relationships.

\subsection{3-D LDM}

The LDM of this research requires the CCPA data as its training target and the GEFS reforecasts as conditional inputs. The training of LDM relies on both the pre-trained VQ-VAE and the 3-D ViT. The LDM was initially validated by computing the loss of reconstructed noise. For fine-tuning, it was later validated by generating forecasts on randomly selected initializations and CRPS as the metric. The hyperparameters of the LDM are summarized in Tabel \ref{tab3}.

\begin{table}[h]
\caption{Identified LDM hyperparameters and their initial search options.}\label{tab3}
\vspace{1ex}
\centering
\begin{tabularx}{\textwidth}{llC}
\\[-1.8ex]\hline 
\hline \\[-1.8ex]
Name & Hyperparameter options & Choice \\
\hline
\multirow{2}*{Diffusion steps} & (1) 50 & \multirow{2}*{(2)} \\
 & (2) 100 & \\
 
\multirow{2}*{Scheduled signal rates} & [0.02, 0.95] & \multirow{2}*{(1)} \\
 & [0.10, 0.95] & \\
 
\multirow{3}*{Hidden layer dimensions} & (1) [32, 64, 96, 128] & \multirow{3}*{(1)} \\
 & (2) [64, 64, 64, 64] & \\
 & (3) [32, 64, 128, 256] & \\

\multirow{2}*{Exponential moving average} & (1) On & \multirow{2}*{(1)} \\
 & (2) Off & \\

\multirow{2}*{Learning rate schedule} & (1) 1E-4 with cosine annealing & \multirow{2}*{(1)} \\
 & (2) Fixed 1E-5 & \\
\hline
\end{tabularx}
\end{table}

\noindent
Linear diffusion schedules were applied in this research. Under this meta, more diffusion steps are generally helpful but also require more computation for sample generations. We are aware that more advanced diffusion model designs and diffusion scheduling have been proposed, and we hope that future studies can introduce them to the field of weather forecasting.

%%%%%%%%%%%%%%%%%%%%%%%%%%%%%%%%%%%%%%%%%%%%%%%%%%%%%%%%%%%%%%%%%%%%%
% REFERENCES
%%%%%%%%%%%%%%%%%%%%%%%%%%%%%%%%%%%%%%%%%%%%%%%%%%%%%%%%%%%%%%%%%%%%%
% Make your BibTeX bibliography by using these commands:
\bibliographystyle{ametsocV6}
\bibliography{references}

\end{document}